\documentclass[a4paper,twocolumn,pre,floatfix,showpacs,aps,10pt]{revtex4-1}

\usepackage{amsfonts,amsmath,amssymb}
\usepackage{textcomp}
\usepackage{bm}
\usepackage{graphicx}
\usepackage[latin1]{inputenc}
\usepackage{rotating}
\usepackage{color}
\usepackage{hyperref}

\newcommand{\ie}{{i.e.}}
\newcommand{\eg}{{e.g.}}
\newcommand{\etal}{\textit{et~al.}}

\newcommand{\notop}{{{}_{}}}
\renewcommand{\vec}[1]{\bm{#1}}

\newcommand{\ii}{\mathrm{i}}

\newcommand{\re}{\mathrm{Re}}

\newcommand{\pp}{\partial^{{}}}

\newcommand{\aO}{a^{{}}_0}
\newcommand{\aOsqr}{a^{2_{}}_0}
\newcommand{\Dth}{D^{{}}_\mathrm{th}}
\newcommand{\Eac}{E_\mathrm{ac}}

\newcommand{\FFFrad}{\vec{F}^\mathrm{rad}}
\newcommand{\Frad}{F^{\mathrm{rad}_{}}}
\newcommand{\pa}{p^\notop_\mathrm{a}}
\newcommand{\pasqr}{p^{2_{}}_\mathrm{a}}
\newcommand{\pI}{p^\notop_1}

\newcommand{\rrr}{\vec{r}}

\newcommand{\UUU}{\vec{U}}
\newcommand{\va}{v^\notop_\mathrm{a}}
\newcommand{\vasqr}{v^{2_{}}_\mathrm{a}}

\newcommand{\fI}{f_1}
\newcommand{\fII}{f_2}

\newcommand{\upvec}{\vec{u}^\mathrm{p}}
\newcommand{\upvecstar}{\vec{u}^\mathrm{p*}}
\newcommand{\up}{u^\mathrm{p}}
\newcommand{\upstar}{u^\mathrm{p*}}
\newcommand{\uradvec}{\vec{u}^\mathrm{rad}}

\newcommand{\urad}{u^\mathrm{rad}}
\newcommand{\ustr}{u^\mathrm{str}}
\newcommand{\ua}{u_\mathrm{a}}
\newcommand{\uO}{u^{{}}_0}
\newcommand{\uarad}{u_\mathrm{a}^\mathrm{rad}}
\newcommand{\uastr}{u_\mathrm{a}^\mathrm{str}}
\newcommand{\ac}{a_\mathrm{c}}

\newcommand{\cO}{c^\notop_0}

\newcommand{\cP}{c^\notop_\mathrm{p}}

\newcommand{\rhoO}{\rho^\notop_0}
\newcommand{\rhoP}{\rho^\notop_\mathrm{p}}
\newcommand{\rhoTi}{\tilde{\rho}}

\newcommand{\sigmaP}{\sigma^\notop_\mathrm{p}}

\newcommand{\deltaTi}{\tilde{\delta}}

\newcommand{\KO}{\kappa^\notop_0}
\newcommand{\KP}{\kappa^\notop_\mathrm{p}}
\newcommand{\kapTi}{\tilde{\kappa}}

\newcommand{\Upp}{U_\mathrm{pp}}

\newcommand{\SIC}{^\circ\!\textrm{C}}

\newcommand{\SIMHz}{\textrm{MHz}}

\newcommand{\SIJpcm}{\textrm{J}\:\textrm{m$^{-3}$}}

\newcommand{\SIkgpcm}{\textrm{kg}\:\textrm{m$^{-3}$}}
\newcommand{\SIm}{\textrm{m}}

\newcommand{\SImm}{\textrm{mm}}
\newcommand{\SImum}{\textrm{\textmu{}m}}

\newcommand{\SImps}{\SIm\,\SIs^{-1}}
\newcommand{\SImumps}{\textrm{\textmu{}m}\,\SIs^{-1}}
\newcommand{\SImmps}{\textrm{mm}\,\SIs^{-1}}

\newcommand{\SImPas}{\textrm{mPa}\:\textrm{s}}

\newcommand{\SIs}{\textrm{s}}

\newcommand{\SImus}{\textrm{\textmu{}s}}

\newcommand{\SIV}{\textrm{V}}

\newcommand{\beq}[1]{\begin{equation} \eqlab{#1}}
\newcommand{\eeq}{\end{equation}}
\newcommand{\bsub}{\begin{subequations}}
\newcommand{\esub}{\end{subequations}}
\def\bal#1\eal{\begin{align}#1\end{align}}
\def\bsubal#1\esubal{\bsub \begin{align}#1\end{align} \esub}

\newcommand{\eqlab}[1]{\label{eq:#1}}
\renewcommand{\eqref}[1]{Eq.~(\ref{eq:#1})}
\newcommand{\eqsref}[2]{Eqs.~(\ref{eq:#1}) and~(\ref{eq:#2})}
\newcommand{\figref}[1]{Fig.~\ref{fig:#1}}
\newcommand{\figsref}[2]{Figs.~\ref{fig:#1} and~\ref{fig:#2}}
\newcommand{\figlab}[1]{\label{fig:#1}}

\newcommand{\secref}[1]{Section~\ref{sec:#1}}
\newcommand{\secsref}[2]{Sections~\ref{sec:#1} and~\ref{sec:#2}}
\newcommand{\seclab}[1]{\label{sec:#1}}
\newcommand{\tabref}[1]{Table~\ref{tab:#1}}

\newcommand{\tablab}[1]{\label{tab:#1}}

\begin{document}

\date{31 August 2012}

\title{Acoustic radiation- and streaming-induced microparticle velocities determined by micro-PIV in an ultrasound symmetry plane}

\author{Rune Barnkob$^{\mathrm{(a)*}}$,
Per Augustsson$^{\mathrm{(b)}}$,
Thomas Laurell$^{\mathrm{(b,c)}}$, and
Henrik Bruus$^{\mathrm{(d)}}$}

\affiliation{%
$^{\mathrm{(a)}}$Department of Micro- and Nanotechnology, Technical University of Denmark\\
DTU Nanotech Building 345 East, DK-2800 Kongens Lyngby, Denmark\\
$^{\mathrm{(b)}}$Department of Measurement Technology and Industrial Electrical Engineering, Division of Nanobiotechnology, Lund University, Box 118, S-221 00 Lund, Sweden\\
$^{\mathrm{(c)}}$Department of Biomedical Engineering, Dongguk University, Seoul, South Korea\\
$^{\mathrm{(d)}}$Department of Physics, Technical University of Denmark\\
DTU Physics Building 309, DK-2800 Kongens Lyngby, Denmark}

\begin{abstract}

We present micro-PIV measurements of suspended microparticles of diameters from $0.6~\SImum$ to $10~\SImum$ undergoing acoustophoresis in an ultrasound symmetry plane in a microchannel. The motion of the smallest particles are dominated by the Stokes drag from the induced acoustic streaming flow, while the motion of the largest particles are dominated by the acoustic radiation force. For all particle sizes we predict theoretically how much of the particle velocity is due to radiation and streaming, respectively. These predictions include corrections for particle-wall interactions and ultrasonic thermoviscous effects, and they match our measurements within the experimental uncertainty. Finally, we predict theoretically and confirm experimentally that the ratio between the acoustic radiation- and streaming-induced particle velocities is proportional to the square of the particle size,
the actuation frequency and the acoustic contrast factor, while it is inversely proportional to the kinematic viscosity.

\pacs{43.20.Ks, 43.25.Qp, 43.25.Nm, 47.15.-x, 47.35.Rs\\\\$^*$ Corresponding author: barnkob@alumni.dtu.dk}
\end{abstract}

%

\maketitle

\section{Introduction}
\label{sec:Introduction}

Acoustofluidics and ultrasound handling of particle suspensions, recently reviewed in \emph{Review of Modern Physics}  \cite{Friend2011} and \emph{Lab on a Chip}  \cite{Bruus2011c}, is a field in rapid growth for its use in biological applications, such as separation and manipulation of cells and bioparticles. Microchannel acoustophoresis has largely been limited to manipulation of micrometer-sized particles, such as yeast \cite{Hawkes2004}, blood cells  \cite{Petersson2004}, cancer cells  \cite{Thevoz2010, Augustsson2012b, Ding2012}, natural killer cells  \cite{Vanherberghen2010}, and affinity ligand complexed microbeads  \cite{Augustsson2012a}, for which the acoustic radiation force dominates. Precise acoustic control of sub-micrometer particles, \eg\ small bacteria, vira, and large biomolecules remains a challenge, due to induction of acoustic streaming of the suspending fluid. Nevertheless, acoustic streaming has been used to enhance the convective transport of substrate in a microenzyme reactor for improved efficiency \cite{Bengtsson2004}, while acoustic manipulation of sub-micrometer particles has been achieved in a few specific cases including enhanced biosensor readout of bacteria \cite{Kuznetsova2005} and bacterial spores  \cite{Martin2005}, and trapping of E-coli bacteria \cite{Hammarstrom2012}.

When a standing ultrasound wave is imposed in a microchannel containing an aqueous suspension of particles, two forces of acoustic origin act on the particles: the Stokes drag force from the induced acoustic streaming and the acoustic radiation force from sound wave scattering on the particles. To date, the experimental work on acoustophoresis has primarily dealt with cases where the acoustic radiation force dominates the motion, typically for particles of diameters larger than $2~\SImum$. Quantitative experiments of 5-$\SImum$-diameter polymer particles in water  \cite{Barnkob2010, Augustsson2011, Barnkob2012} have shown good agreement with the classical theoretical predictions \cite{Yosioka1955, Gorkov1962} of the acoustic radiation force acting on a microparticle of radius $a$ much smaller than the acoustic wavelength $\lambda$, and where the viscosity of the suspending fluid is neglected. However, as the particle diameter is decreased below $2~\SImum$, a few times the acoustic boundary-layer thickness, the particle motion is typically strongly influenced by the Stokes drag force from the induced acoustic streaming flow, which has been reported by several groups \cite{Wiklund2012, spengler2003, Hagsater2007}, and the radiation force is modified due to the acoustic boundary layer  \cite{Settnes2012}.

As pointed out in a recent review  \cite{Wiklund2012}, the acoustic streaming is difficult to fully characterize due to its many driving mechanisms and forms. In acoustofluidic systems, the streaming is primarily boundary-driven arising at rigid walls from the large viscous stresses inside the sub-micrometer-thin acoustic boundary layer of width $\delta$. The boundary-driven acoustic streaming was theoretically treated by Rayleigh  \cite{LordRayleigh1884} for an isothermal fluid in an infinite parallel-plate channel with a standing wave parallel to the plates of wavelength $\lambda$ much larger than the plate distance $h$, and where $h$ is large compared to $\delta$, \ie\ $\lambda\gg h\gg\delta$. However, in many applications of acoustofluidic systems the channels provide enhanced confinement, the acoustic wavelength is comparable to the channel height, and the liquid cannot be treated as being isothermal. Rayleigh's prediction is often cited, but to our knowledge the literature contains no quantitative validation of its accuracy when applied to acoustofludic systems. This lack of quantitative tests is most likely due to the fact that boundary-driven acoustic streaming is very sensitive to geometry and boundary conditions, making it difficult to achieve sufficient experimental control. However, quantitative comparisons between theory and experiment of acoustic streaming are crucial for the advancement of the acoustofluidics research field. Understanding and controlling the ratio of radiation- and streaming-induced acoustophoretic velocities may be the key for future realization of ultrasound manipulation of sub-micrometer particles.

In 2011, we presented a temperature-controlled micro-PIV setup for accurate measurements of the acoustophoretic microparticle motion in a plane \cite{Augustsson2011}. Here, we use the same system and the ability to establish a well-controlled transverse resonance for quantitative studies of how much the radiation- and streaming-induced velocities, respectively, contribute to the total acoustophoretic velocity. More specifically, as illustrated in \figref{chip_sketch}, we study the microparticle motion in the ultrasound symmetry plane (magenta) of a straight rectangular microchannel of width $w = 377~\SImum$ and height $h = 157~\SImum$. We determine the velocities for particles of diameter $2a$ ranging from 0.6~$\SImum$ to 10~$\SImum$, and based on this we examine the validity of Rayleigh's theoretical streaming prediction. We also derive theoretically and validate experimentally an expression for the microparticle velocity as function of particle size, ultrasound frequency, and mechanical properties of the suspending medium.

\section{Theory of single-particle acoustophoresis}
\seclab{SingleParticle}

In this work we study a silicon-glass chip containing a rectangular microchannel sketched in \figref{chip_sketch} and described further in \secref{experiment}. The microchannel contains a particle suspension, and the chip is ultrasonically actuated by attaching a piezo transducer to the chip and driving it with the voltage $\Upp$ at the angular frequency $\omega=2\pi f$, where $f$ is a frequency in the low MHz range. By proper tuning of the applied frequency, the actuation induces a resonant time-harmonic ultrasonic pressure field $p_1(\vec{r})\exp{(-\ii\omega t)}$ and velocity field $\vec{v}_1(\vec{r})\exp{(-\ii\omega t)}$, here expressed in the complex time-harmonic notation. Throughout the paper, we only study the case of a 1D transverse pressure resonance of amplitude $\pa$ and wavenumber $k = 2\pi/\lambda$,
 \beq{p1D}
 \pI(\rrr) = \pa\:\cos\Big[k\Big(y - \frac{w}{2}\Big)\Big].
 \eeq
The case of $\lambda/2 = w$ or $k = \pi/w$ is shown in \figref{symmetry_plane}(a).

The particle suspensions are dilute enough that the particle-particle interactions are negligible, and thus only single-particle effects are relevant. These comprise the acoustic radiation force due to particle-wave scattering and the viscous Stokes drag force from the acoustic streaming flow. Both effects are time-averaged second-order effects arising from products of the first-order fields. The drag force from the acoustic streaming flow dominates the motion of small particles, while the motion of larger particles are dominated by the acoustic radiation force. This is clearly illustrated in recent numerical simulations by Muller \etal\ \cite{Muller2012}, which are reproduced in \figref{symmetry_plane}: (b) the streaming flow advects small particles in a vortex pattern, and (c) radiation force pushes larger particles to the pressure nodal plane at $y=0$.

\begin{figure}[t]
  \centering
  \includegraphics[width=85mm]{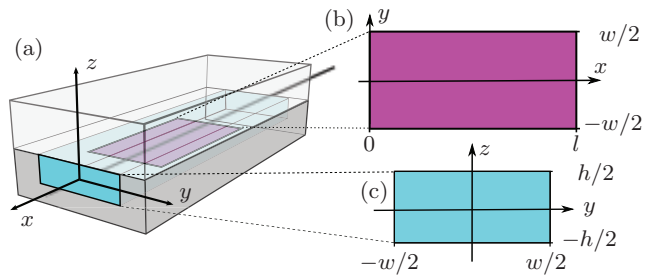}
\caption{\figlab{chip_sketch} (a) Sketch of the silicon/glass microchip used in our experiments, see also Ref.~\cite{Augustsson2011}. It contains a straight rectangular water-filled microchannel (light cyan) of length $L = 35~\SImm$, width $w=377~\SImum$, and  height $h=157~\SImum$. (b) The horizontal ultrasound symmetry plane (magenta) of length $l=892~\SImum$ and width $w$ in the $xy$ plane at the center of the channel. (c) The vertical channel cross section (cyan).}
\end{figure}

\begin{figure*}
  \centering
  \includegraphics[width=170mm]{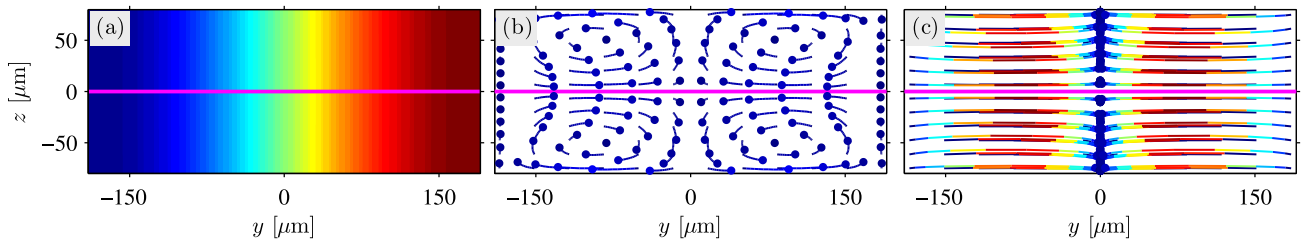}
\caption{\figlab{symmetry_plane} Numerical simulation of microparticle acoustophoresis in the vertical microchannel cross section of \figref{chip_sketch}(c) adapted from Ref.~\cite{Muller2012}. The magenta line represents the ultrasound symmetry plane of \figref{chip_sketch}(b). (a) Color plot of the transverse standing ultrasound pressure wave $p_1$ of \eqref{p1D} ranging from $-p_a$ (dark blue) to $p_a$ (dark red). (b) Trajectories (blue lines) of small 0.5-$\SImum$-diameter particles (dots) dominated by the Stokes drag force from the boundary-induced streaming. (c) Trajectories (colored lines) of large 5.0-$\SImum$-diameter particles (dots) dominated by the acoustic radiation force.}
\end{figure*}

\subsection{The acoustic radiation force}
\seclab{Frad}

We consider a spherical particle of radius $a$, density $\rhoP$, and compressibility $\KP$ suspended in a liquid of density $\rhoO$, compressibility $\KO$, viscosity $\eta$, and momentum diffusivity $\nu=\eta/\rhoO$. Recently, Bruus and Settnes  \cite{Settnes2012} gave an analytical expression for the viscosity-dependent time-averaged radiation force $\FFFrad$ in the experimentally relevant limit of the wavelength $\lambda$ being much larger than both the particle radius $a$ and the thickness $\delta =\sqrt{2\nu/\omega}$ of the acoustic boundary layer, without any restrictions on the ratio $\deltaTi = \delta/a$. For the case of a 1D transverse pressure resonance, \eqref{p1D}, the viscosity-dependent acoustic radiation force on a particle located at $(y,z)$ reduces to the $z$-independent expression
 \beq{Frad1D}
 \Frad(y,z) = 4\pi a^3 k\Eac\:\Phi(\kapTi,\rhoTi,\deltaTi)\:\sin\big(2k y\big),
 \eeq
where $\Eac=\tfrac{1}{4}\KO\pasqr$ is the time-averaged acoustic energy density and	where the acoustic contrast factor $\Phi$ is given in terms of the material parameters as
 \bsub
 \begin{alignat}{2}
 \eqlab{PhiDef}
 \Phi(\kapTi,\rhoTi,\deltaTi) &=
 \frac{1}{3}\fI(\kapTi)  \!+\! \frac{1}{2}\re\big[\fII(\rhoTi,\deltaTi)\big],&\qquad&\\
 \eqlab{f1RES}
 \fI(\kapTi) &=  1 - \kapTi,
 &  \text{ with }
 \kapTi &= \frac{\KP}{\KO},\\
 \eqlab{f2Res}
 \fII(\rhoTi,\deltaTi) &= \frac{2\big[1\!-\!\Gamma(\deltaTi)\big]
 (\rhoTi\!-\!1)}{2\rhoTi+1-3\Gamma(\deltaTi)},
 & \text{ with }
 \rhoTi &= \frac{\rhoP}{\rhoO},\\
 \eqlab{gamRES}
 \Gamma(\deltaTi) &=
 -\frac{3}{2} \Big[1+\ii(1+\deltaTi)\Big] \deltaTi,
 & \text{ with }
 \deltaTi &= \frac{\delta}{a}.
\end{alignat}
\esub
We note that for all the microparticle suspensions studied in this work including the viscous 0.75:0.25 water:glycerol mixture, the viscous corrections to $\Phi$ are negligible as we find $\big|\Phi(\kapTi,\rhoTi,\deltaTi)/\Phi(\kapTi,\rhoTi,0)-1 \big| < 0.4~\%$.

If $\FFFrad$ is the only force acting on a suspended particle, the terminal speed of the particle is ideally given by the Stokes drag as $\uradvec = \FFFrad/(6\pi\eta a)$. Using \eqref{Frad1D} for the transverse resonance, $\uradvec$ only has a horizontal component $\urad_y$, and this can be written in the form
 \beq{urady}
 \urad_y = \uO\: \frac{a^2}{\aOsqr}\:\sin(2ky),
 \eeq
where the characteristic velocity amplitude $\uO$ and particle radius $\aO$ are given by
 \bsub
 \eqlab{u0_a0}
 \begin{alignat}{3}
 \uO &= \frac{4\Eac}{\rhoO\cO}
 &&= \frac{4\Eac}{Z_0} &&= 27~\SImumps,\\
 \aO &= \sqrt{\frac{6\nu}{\Phi}\frac{1}{\omega}}
 &&= \delta\sqrt{\frac{3}{\Phi}} &&= 1.6~\SImum.
 \end{alignat}
 \esub
Here, $Z_0$ is the characteristic acoustic impedance, and the numerical values are calculated for polystyrene particles suspended in water using parameter values listed in \secref{experiment} with  $f = 2~\SIMHz$ and $\Eac=10~\SIJpcm$ as in Barnkob \textit{et al.} \cite{Barnkob2010, Barnkob2012}

\subsection{Boundary-driven acoustic streaming}

In 1884 Lord Rayleigh  \cite{LordRayleigh1884} published his now classical analysis of the boundary-driven acoustic streaming velocity field $\UUU$ in an infinite parallel-plate channel induced by a first-order bulk velocity field having only a horizontal $y$-component given by $v_1 = \va\sin\big[k(y-w/2)\big]$. This corresponds to the first-order pressure of \eqref{p1D} illustrated in \figref{symmetry_plane}(a). For an isothermal fluid in the case of  $\lambda \gg h \gg \delta$, Rayleigh found the components $U_y$ and $U_z$ of $\UUU$ outside the acoustic boundary to be
 \bsub
 \eqlab{Rayleigh}
 \bal
 U_y(y,z) &= \frac{3}{8}\frac{\vasqr}{\cO}\sin{(2ky)}\Bigg[
 \frac{3}{2}\frac{z^2}{(h/2)^2} - \frac{1}{2}\Bigg],\\
 U_z(y,z) &= \frac{3}{16}\frac{\vasqr}{\cO}kh\cos{(2ky)}\Bigg[
 \frac{z}{(h/2)} - \frac{z^3}{(h/2)^3}\Bigg],\\
 \va &= \frac{\pa}{\rhoO\cO} = 2 \sqrt{\frac{\Eac}{\rhoO}}.
 \eal
 \esub
A plot of $\UUU$ driven by the 1D transverse standing half-wave resonance is shown in \figref{ustr}. We expect this analytical expression to deviate from our measurements because the actual channel does have side walls, it is not isothermal, and instead of $\lambda \gg h$ we have $\lambda = 4.8 h$ for $\lambda = 2w$ and $\lambda = 2.4h$ for $\lambda = w$.

At the ultrasound symmetry plane $z=0$, $\UUU$ only has a horizontal component, which we denote $\ustr_y$. In analogy with \eqref{urady} this can be written as
 \beq{ustr1D}
 \ustr_y = \uO\: s^0_\mathrm{p}\:\sin(2ky), \quad
 s^0_\mathrm{p} = \frac{3}{16} \approx 0.188,
 \eeq
where the sub- and superscript in the streaming coefficient $s^0_\mathrm{p}$ refer respectively to the parallel-plate geometry and the isothermal liquid in Rayleigh's analysis.

To estimate the effect on $\ustr_y$ of the side walls and the large height $h \approx \lambda$  in the rectangular channel of \figref{chip_sketch}(c), we use the numerical scheme developed by Muller \etal\ \cite{Muller2012} for calculating the acoustic streaming based directly on the hydrodynamic equations and resolving the acoustic boundary layers, but without taking thermoviscous effects fully into account. The result shown in \figref{ustr_along_y} reveals that $\ustr_y$ is suppressed by a factor of 0.82 in the rectangular geometry relative to the parallel-plate geometry and that it approaches zero faster near the side walls at $y = \pm w/2$. The approximate result is
 \beq{ustr1Dr}
 \ustr_y \approx \uO\: s^0_\mathrm{r}\:\sin(2ky), \quad
 s^0_\mathrm{r} \approx 0.154,
 \eeq
where the sub- and superscript in the streaming coefficient $s^0_\mathrm{r}$ refer respectively to the rectangular geometry and the isothermal liquid.

\begin{figure}[!t]
  \centering
    \includegraphics[width=80mm]{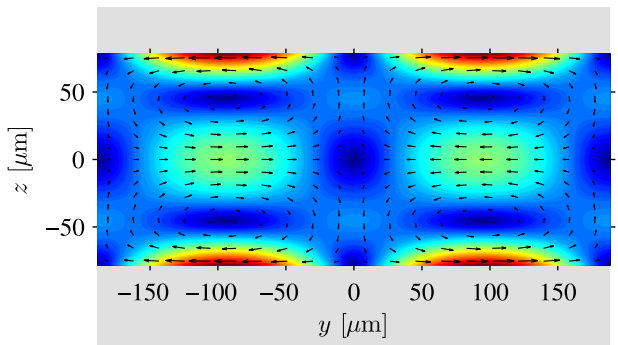}
	\caption{\figlab{ustr}
Vector plot (black arrows) of the acoustic streaming $\vec{U}$ and color plot of its magnitude $U$ from zero (dark blue) to maximum (dark red) given by Rayleigh's analytical expression \eqref{Rayleigh} and valid for a shallow isothermal liquid slab ($\lambda \gg h$) between two parallel plates (gray) of distance $h$ driven by a standing ultrasound pressure wave $\pI=\pa\cos\big[k(y-w/2)\big]$ for $\lambda/2 = w$ or $k = \pi/w$ with $w = 377~\SImum$.}
\end{figure}

We estimate the thermoviscous effect on $\ustr_y$, in particular the temperature dependence of viscosity, using the analytical result by Rednikov and Sadhal for the parallel-plate geometry \cite{Rednikov2011}. They found a streaming factor $s^T_\mathrm{p}$ enhanced relative to $s^0_\mathrm{p}$,
 \bsubal
 s^T_\mathrm{p} &= \Big(1+\frac{2}{3}B^\notop_T\Big)\: s^0_\mathrm{p}
 \approx 1.26\: s^0_\mathrm{p}, \\
 B^\notop_T &=  (\gamma-1)\Bigg[1- \frac{\big(\pp_T\eta\big)^{{}}_p}{\eta\alpha}
 \Bigg] \frac{\sqrt{\nu \Dth}}{\nu + \Dth},
 \esubal
where $\alpha$ is the thermal expansion coefficient, $\Dth$ the thermal diffusivity, and $\gamma$ the specific heat ratio, and where the value is calculated for water at $T = 25~\SIC$.

Combining the reduction factor 0.82 from the rectangular geometry with the enhancement factor 1.26 from thermoviscous effects leads to $s^T_\mathrm{r} \approx 1.03\: s^0_\mathrm{p}$ or
 \beq{ustr1DT}
 \ustr_y \approx \uO\: s^T_\mathrm{r}\:\sin(2ky), \quad
 s^T_\mathrm{r} \approx 0.194,
 \eeq
where the sub- and superscript in the streaming coefficient $s^T_\mathrm{r}$ refer respectively to the rectangular geometry and a thermoviscous liquid, see \figref{ustr_along_y}.

\begin{figure}[!t]
  \centering
  \includegraphics[width=80mm]{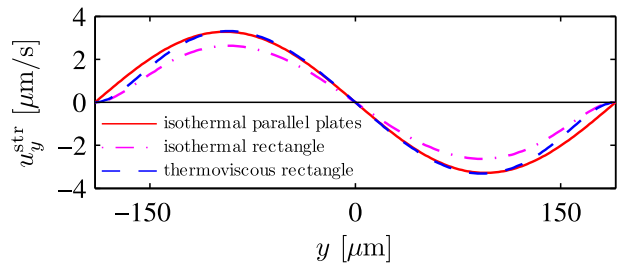}
\caption{\figlab{ustr_along_y} Numerical simulation of the horizontal component $\ustr_y$  of the streaming velocity in the ultrasound symmetry plane at $z=0$. In this plane the vertical component is zero, $\ustr_z = 0$. Three cases are shown: the isothermal parallel-plate channel \eqref{ustr1D}, the isothermal rectangular channel \eqref{ustr1Dr}, and the thermoviscous rectangular channel \eqref{ustr1DT}.}
\end{figure}

\subsection{Acoustophoretic particle velocity}

\begin{figure*}[!t]
  \centering
  \includegraphics[]{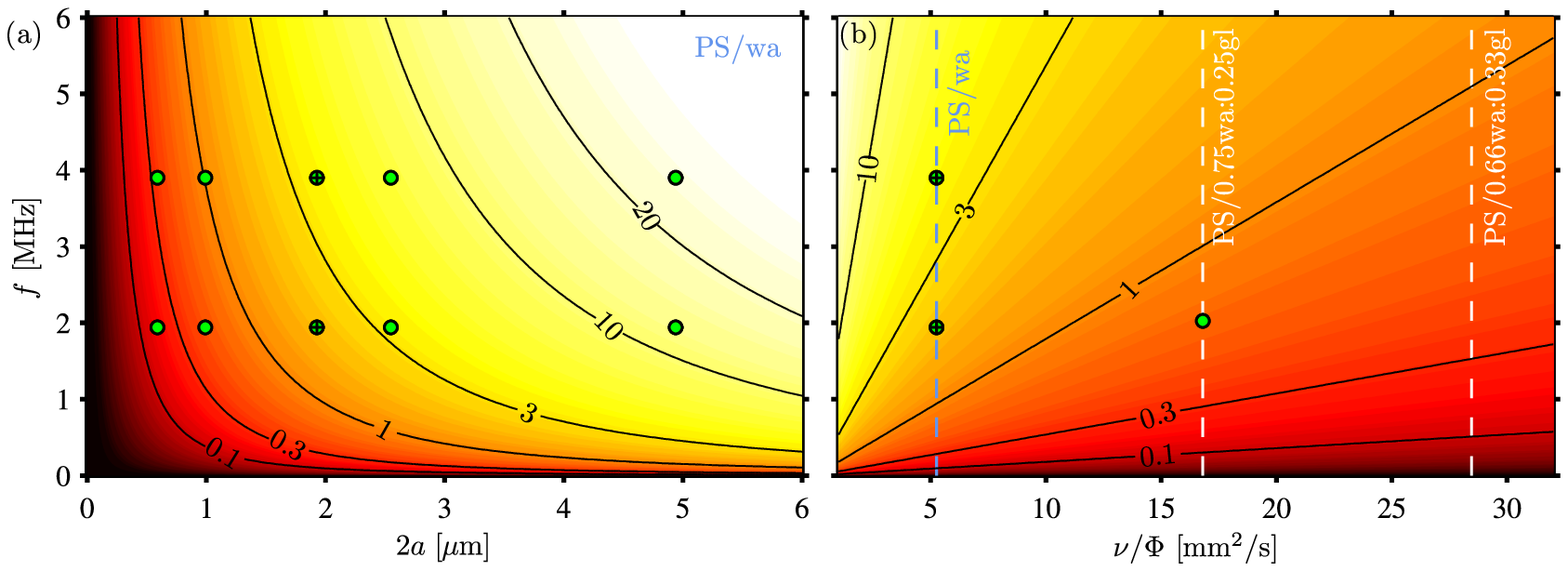}
\caption{\figlab{ratio_vs_f_nuPhi_asqr} Theoretical prediction of the ratio $\urad_y/\ustr_y$ \eqref{ratio} of radiation- and streaming-induced velocities in the ultrasound symmetry plane of the microchannel (magenta in \figsref{chip_sketch}{symmetry_plane}) at 25~$\SIC$. (a) Contour plot of $\urad_y/\ustr_y$ as function of particle diameter $2a$ and ultrasound frequency $f$ for a suspension of polystyrene particles PS in pure water. (b) Contour plot of $\urad_y/\ustr_y$ as function of $f$ and the rescaled momentum diffusivity $\nu/\Phi$ for fixed particle diameter $2a=2~\SImum$. The green dots indicate the cases studied experimentally in \secsref{experiment}{results}. Suspensions of polystyrene particles in three different liquids are indicated by dashed lines: pure water (light blue) as well as  0.75:0.25 and 0.67:0.33 mixtures of water:glycerol (white).}
\end{figure*}

A single particle undergoing acoustophoresis is directly acted upon by the acoustic radiation force $\FFFrad$, while the acoustic streaming of velocity $\vec{U}$ contributes with a force on the particle through the viscous Stokes drag from the suspending liquid. Inertial effect can be neglected as the characteristic time scale $\rhoP a^2/\eta$ of acceleration  $( < 1~\SImus$) is minute in comparison with the time scale of the motion of particles ($> 1$~ms). The equation of motion for a spherical particle of velocity $\upvec$ then becomes
 \beq{upvec}
 \upvec = \frac{\FFFrad}{6\pi\eta a} + \vec{U}.
 \eeq
As we have seen above, there are no vertical velocity components in the ultrasound symmetry plane at $z=0$, and combining \eqsref{urady}{ustr1DT} we obtain the horizontal particle velocity component $\up_y$ of amplitude $\ua$,
  \beq{up}
  \up_y = \urad_y + \ustr_y = \ua\:\sin(2k y), \quad
  \ua = \uO\frac{a^2}{\aOsqr} + \uO\:s,
  \eeq
where we have dropped the sub- and superscripts of the streaming coefficient $s$. The ratio of the radiation- and streaming-induced velocity amplitudes becomes
\beq{ratio}
  \frac{\urad_y}{\ustr_y} = \frac{1}{s}\:\frac{a^2}{\aOsqr}
  = \frac{1}{6s}\frac{\Phi}{\nu}\:\omega a^2,
\eeq
which scales linearly with the angular frequency $\omega$ and the square $a^2$ of the particle radius, but inversely with the streaming coefficient $s$ and the momentum diffusivity $\nu/\Phi$ rescaled by the acoustic contrast factor.

In \figref{ratio_vs_f_nuPhi_asqr} we show colored contour plots of the ratio $\urad_y/\ustr_y$: in (a) for polystyrene particles in water at $25~\SIC$ as function of the particle diameter $2a$ and the ultrasound frequency $f$, and in (b) as function of $f$ and the rescaled momentum diffusivity $\nu/\Phi$ for fixed particle diameter $2a=2~\SImum$. The green dots indicate the experiments described in \secsref{experiment}{results}.

We define the critical particle diameter $2\ac$ for cross-over from radiation-dominated to streaming-dominated acoustophoresis as the particle diameter for which $\urad_y(\ac) = \ustr_y(\ac)$. This results in
\bal\label{eq:ac}
	2\ac = 2a_0\,\sqrt{s} = \sqrt{\frac{24\nu}{\omega}\frac{s}{\Phi}} \approx 1.4~\SImum,
\eal
where the numerical value is calculated for polystyrene particles in water ($\Phi=0.17$) at $f = 2$~MHz using $s = s^T_\mathrm{r}$. For $a=\ac$ the ratio of the velocity amplitudes $\urad_y/\ustr_y$ is unity, and consequently the unity contour line in \figref{ratio_vs_f_nuPhi_asqr}(a) represents $2\ac$ as function of ultrasound frequency $f$.

\begin{table}[!t]
\caption{\tablab{wall_corr} The wall correction factor $\chi$ to the single-particle drag for the particle sizes used in the experiment.}
\centering
\begin{ruledtabular}
\begin{tabular}{cccc}
2a & $\chi^\mathrm{paral}_{z=0}$ & $\chi^\mathrm{paral}_{z=\pm h/4}$
   & $\chi^\mathrm{perp}_{\Delta y=w/4}$ \\\hline
$0.6~\SImum$ &   1.004 & 1.005 & 1.004 \\
$1.0~\SImum$ &   1.006 & 1.008 & 1.006 \\
$1.9~\SImum$ &   1.012 & 1.016 & 1.011 \\
$2.6~\SImum$ &   1.017 & 1.022 & 1.016 \\
$4.9~\SImum$ &   1.032 & 1.042 & 1.030 \\
$10.2~\SImum$ &  1.070 & 1.092 & 1.065 \\
\end{tabular}
\end{ruledtabular}
\end{table}

\subsection{Wall corrections to single-particle drag}
\seclab{wall_effects}

The sub-millimeter width and height of the rectangular microchannel enhance the hydrodynamic drag on the microparticles. This problem was treated by Fax\'{e}n for of a sphere moving parallel to a planar wall or in between a pair of parallel planar walls  \cite{Faxen1922} and later extended by Brenner \cite{Brenner1961} to motion perpendicular to a single planar wall, as summarized by Happel and Brenner  \cite{Happel1983}. The enhancement of the Stokes drag is characterized by a dimensionless correction factor $\chi(a)$ modifying  \eqref{up},
  \beq{upvecwall}
  \up_y = \bigg[\frac{1}{\chi(a)}\:\frac{a^2}{\aOsqr} + s \bigg] \uO \sin(2k y).
  \eeq

No general analytical form exists for $\chi$, so we list the result for three specific cases.
For a particle moving parallel to the surface in the symmetry plane $z=0$
in the gap of height $h$ between two parallel planar walls, $\chi$ is
 \beq{gamma_parallel_half}
 \chi_{z=0}^\mathrm{paral}
 \approx \big[1-1.004(2a/h)+0.418(2a/h)^3\big]^{-1}\approx 1.070,
 \eeq
while for motion in the planes at $z=\pm h/4$ it is
 \beq{gamma_parallel_quarter}
 \chi_{z=\pm h/4}^\mathrm{paral}
 \approx \big[1-1.305(2a/h)+1.18(2a/h)^3\big]^{-1}\approx 1.092.
 \eeq
Here the numerical values refer to a particle with diameter $2a = 10~\SImum$ moving in a gap of height $h = 157~\SImum$. Similarly, for particle motion perpendicular to a single planar wall, the correction factor is
 \bal\eqlab{gamma_perp}
 \chi^\mathrm{perp} &= \frac{4}{3}\sinh{(\alpha)}\sum_{n=1}^{\infty}
 \frac{n(n+1)}{(2n-1)(2n+3)}
 \nonumber\\ &\quad\times \bigg[\frac{2\sinh{[(2n+1)\alpha]}+(2n+1)\sinh{(2\alpha)}}{
 4\sinh^2{[(n+\tfrac{1}{2})\alpha]}-(2n+1)^2\sinh^2{(\alpha)}}-1\bigg]
 \nonumber \\ &\approx 1.065,
 \eal
where $\alpha=\cosh^{-1}(\Delta y/a)$ and $\Delta y$ is the distance from the center of the particle to the wall. The numerical value refers to a 10-$\SImum$ particle located at $\Delta y=w/4$.

The values of the wall correction factor $\chi$ for all the particle sizes used in this work are summarized in \tabref{wall_corr}.

\section{Experimental procedure}
\seclab{experiment}

Experiments were carried out to test the validity of the theoretical predictions for the acoustophoretic particle velocity \eqref{upvecwall} in the horizontal ultrasound symmetry plane and for the ratio of the corresponding radiation and streaming-induced velocities, see \eqref{ratio} and \figref{ratio_vs_f_nuPhi_asqr}.

\begin{table}[!t]
\caption{\tablab{parameters} Material parameters at ${T=25~\SIC}$. }
\centering
\begin{ruledtabular}
\begin{tabular}{lcc}
 \multicolumn{3}{c}{Polystyrene} \\ \hline
 Density \cite{crc} & $\rhoP$ &\;\,$1050~\SIkgpcm$\\
 Speed of sound \cite{Bergmann1954} (at 20~$\SIC$) & $\cP$ & $2350~\SImps$\\
 Poisson's ratio \cite{Mott2008} & $\sigmaP$ & $0.35$\\
 Compressibility\footnote{Calculated as $\KP=\frac{3(1-\sigmaP)}{1+\sigmaP}\frac{1}{(\rhoP c_\mathrm{p}^2)}$ from Ref.~\cite{Landau1986}.} & $\KP$ & $249~\mathrm{TPa}^{-1}$\\\hline\hline
 \multicolumn{3}{c}{Water} \\ \hline
 Density \cite{crc} & $\rho_0$ &\;\,$997~\SIkgpcm$\\
 Speed of sound \cite{crc} & $c_0$ & $1497~\SImps$\\
 Viscosity \cite{crc} & $\eta$ & $0.890~\SImPas$\\
 Viscous boundary layer, 1.940~MHz & $\delta$ & $0.38~\SImum$ \\
 Viscous boundary layer, 3.900~MHz & $\delta$ & $0.27~\SImum$ \\
 Compressibility\footnote{Calculated as $\KO=1/(\rhoO c_0^2)$} & $\KO$ & $ 448~\mathrm{TPa}^{-1}$\\
 Compressibility factor (polystyrene) & $f_1$ & $0.444$\\
 Density factor (polystyrene) & $f_2$ & $0.034$\\
 Contrast factor (polystyrene) & $\Phi$ & $0.17$\\
 Rescaled momentum diffusivity (polyst.) & $\nu/\Phi$ & $5.25$ mm$^2$\;s$^{-1}$\\\hline\hline
 \multicolumn{3}{c}{0.75:0.25 mixture of water and glycerol} \\ \hline
 Density  \cite{Cheng2008} & $\rho_0$ & $1063~\SIkgpcm$\\
 Speed of sound  \cite{Fergusson1954} & $c_0$ & $1611~\SImps$\\
 Viscosity  \cite{Cheng2008} & $\eta$ & $1.787~\SImPas$\\
 Viscous boundary layer, 2.027~MHz & $\delta$ & $0.51~\SImum$ \\
 Compressibility\footnotemark[2] & $\KO$ & $ 363~\mathrm{TPa}^{-1}$\\
 Compressibility factor (polystyrene) & $f_1$ & $0.313$\\
 Density factor (polystyrene) & $f_2$ & $-0.008$\\
 Contrast factor (polystyrene) & $\Phi$ & $0.10$\\
 Rescaled momentum diffusivity (polyst.) & $\nu/\Phi$ & $16.8$ mm$^2$\;s$^{-1}$\\
\end{tabular}
\end{ruledtabular}
\end{table}

We use the experimental technique and micro-PIV system as presented in Augustsson \etal\ \cite{Augustsson2011}. The setup is automated and temperature controlled. This enables stable and reproducible generation of acoustic resonances as a function of temperature and frequency. It also enables repeated measurements that lead to good statistics in the micro-PIV analyses. The resulting acoustophoretic particle velocities are thus of high precision and accuracy.

Using the chip sketched in \figref{chip_sketch}, a total of 22 sets of repeated velocity measurement cycles were carried out on polystyrene particles of different diameters undergoing acoustophoresis in different suspending liquids and at different ultrasound frequencies. In the beginning of each measurement cycle, a particle suspension was infused in the channel while flushing out any previous suspensions. Subsequently, the flow was stopped, and a time lapse microscope image sequence was recorded at the onset of the ultrasound. The cycle was then repeated.

\subsection{Microparticle suspensions}
\seclab{suspensions}

\begin{table}[!t]
\caption{\tablab{diameter} The nominal and the measured diameter of the polystyrene particles used in the experiment.}
\centering
\begin{ruledtabular}
\begin{tabular}{cc}
Nominal diameter & Measured diameter ($2a$)   	\\ \hline
591~nm       & $(0.59\pm0.03)~\SImum$\footnote{Value from manufacturer and assumed 5~\% standard deviation.}\\
992~nm       & $(0.99\pm0.05)~\SImum$\footnotemark[1]  \\
2.0~$\SImum$ & $(1.91\pm 0.07)~\SImum$\footnote{Measured by Coulter counter} \\
3.0~$\SImum$ & $(2.57\pm 0.07)~\SImum$\footnotemark[2]  \\
5~$\SImum$   & $(5.11\pm 0.16)~\SImum$\footnotemark[2]  \\
10~$\SImum$  &$(10.16\pm 0.20)~\SImum$\footnotemark[2]  \\
\end{tabular}
\end{ruledtabular}
\end{table}

Two types of microparticle suspensions were examined; polystyrene particles suspended in Milli-Q water and polystyrene particles suspended in a 0.75:0.25 mixture of Milli-Q water and glycerol. To each of the two suspending liquids was added 0.01~\% w/V Triton-X surfactant. The material parameters of the suspensions are listed in \tabref{parameters}. Note that the rescaled momentum diffusivity $\nu/\Phi$ of the glycerol suspension is 3 times larger than that for the Milli-Q water suspension.

We analyzed 12 particle suspensions by adding particles of 6 different diameters $2a$ from $0.6~\SImum$ to $10~\SImum$ to the two liquids. The particle diameters were measured using a Coulter Counter (Multisizer 3, Beckman Coulter Inc., Fullerton, CA, USA) and fitting their distributions to Gaussian distributions, see Supplemental Material. The resulting diameters are listed in \tabref{diameter}.

The concentration $C$ of the particles were calculated based on the concentrations provided by the manufacturer and varies in this work from  $10^{10}~\SIm^{-3}$ for the largest particles in the 0.75:0.25 mixture of water and glycerol to $10^{15}~\SIm^{-3}$ for the smallest particles in the pure water solution. The concentrations correspond to mean inter particle distances $C^{-1/3}$ ranging from 4 particle diameters for the largest 10-$\SImum$ particle in water to 173 particle diameters for the smallest 0.6-$\SImum$ particle in the 0.75:0.25 mixture of water and glycerol. Mikkelsen and Bruus \cite{Mikkelsen2005} have reported that hydrodynamic effects become significant for interparticle distances below 2 particle diameters. Thus we can apply the single-particle theory presented in \secref{SingleParticle}.

\subsection{Measurement series}
\seclab{meas_series}

We measured the acoustophoretic velocities of poly\-sty\-rene microparticles in the following four series of experiments, the second being a repeat of the first:\\[2mm]
\textbf{MQ0}: Milli-Q water, $f=1.940~\SIMHz$,\\
\indent $\lambda=2w$, and $2a = 1.0$, 1.9, 2.6, and $5.1~\SImum$.\\
\textbf{MQ1}: Milli-Q water, $f=1.940~\SIMHz$,\\
\indent $\lambda=2w$, and $2a = 0.6$, 1.0, 1.9, 2.6, 5.1, and $10.2~\SImum$.\\
\textbf{MQ2}: Milli-Q water, $f=3.900~\SIMHz$ \\
\indent $\lambda=w$,\hspace*{2mm} and $2a = 0.6$, 1.0, 1.9, 2.6, 5.1, and $10.2~\SImum$.\\
\textbf{GL2}: 0.75:0.25 Milli-Q water:glycerol, $f=2.027~\SIMHz$\\
\indent $\lambda=2w$, and $2a = 0.6$, 1.0, 1.9, 2.6, 5.1, and $10.2~\SImum$.\\[2mm]
Given the different particle diameters, we thus have the above-mentioned 22 sets of acoustophoretic particle-velocity measurements, each consisting of 50 to 250 measurement cycles. All experiments were carried out at a fixed temperature of $25~\SIC$ and the applied piezo voltage $\Upp^*$. The camera frame rate was chosen such that the particles would move at least a particle diameter between two consecutive images. The measurement field of view was $1280\times 640$ pixels corresponding to $892~\SImum\times 446~\SImum$.  The imaging parameters were: optical wavelength 520~nm for which the microscope objective is most sensitive, numerical aperture 0.4, and magnification 20.  See acquisition details in the Supplemental Material.

\subsection{Micro-PIV analysis}
\seclab{microPIV_analysis}

The micro-PIV analyses were carried out using the software \textit{EDPIV - Evaluation Software for Digital Particle Image Velocimetry}, including the image procedure, the averaging in correlation space, and the window shifting described in detail in Ref.~\cite{Augustsson2011}. For the MQ1, MQ2, and Gl1 series, the interrogation window size was 32 $\times$ 32 pixels with a 50~\% overlap resulting in a $79\times 39$ square grid with $16$ pixels between each grid point. For the MQ0 series, the interrogation window size was 64 $\times$ 64 pixels with a 50~\% overlap resulting in a $39\times 19$ square grid with $32$ pixels between each grid point.

In micro-PIV all particles in the volume are illuminated, and the thickness of the measurement plane is therefore related to focal depth of the microscope objective. This thickness, denoted the depth of correlation (DOC), is defined as twice the distance from the measurement plane to the nearest plane for which the particles are sufficiently defocused such that it no longer contributes significantly to the cross-correlation analysis \cite{Meinhart2000}. The first analytical expression for the DOC was derived by Olsen and Adrian \cite{Olsen2000} and later improved by Rossi \etal\ \cite{Rossi2012}. Using the latter, we found that the DOC ranges from $14~\SImum$ to $94~\SImum \approx h/2$ for the smallest and the largest particles, respectively. Consequently, in the vertical direction all observed particles reside within the middle half of the channel.

\section{Results}
\seclab{results}

The core of our results is the 22 discrete acou\-sto\-pho\-retic particle-velocity fields obtained by micro-PIV analysis of the 22 sets of acoustic focusing experiments and shown in the Supplemental Material. As in Ref.~\cite{Augustsson2011}, the measured microparticle velocities $\upvec$ are thus represented on a discrete $x_n\times y_m$ micro-PIV grid
\bal
 \upvec = \upvec(x_n,y_m) = \Bigg[
 \begin{array}{c}
 \up_x(x_n,y_m) \\[2mm]
 \up_y(x_n,y_m)
 \end{array} \Bigg].
 \eal
All measured velocities presented in the following are normalized to their values at $\Upp=1~\SIV$ using the voltage-squared law \cite{Barnkob2010},
\bal\eqlab{Upp_norm_upvec}
  \upvec = \bigg(\frac{1~\SIV}{\Upp^*}\bigg)^2\upvecstar,
\eal
where the asterisk denotes the actual measured values. As a result, the extracted velocity amplitudes and acoustic energy densities are normalized as well,
 \bsubal
 \eqlab{ua_norm}
 \ua &= \bigg(\frac{1~\SIV}{\Upp^*}\bigg)^2\ua^*,\\
 \eqlab{Eac_norm}
 \Eac &= \bigg(\frac{1~\SIV}{\Upp^*}\bigg)^2\Eac^*.
 \esubal
The actual peak-to-peak values of the applied voltage $\Upp^*$ for all four experimental series are given in the Supplemental Material.

\begin{figure}[!t]
  \centering
    \includegraphics[width=85mm]{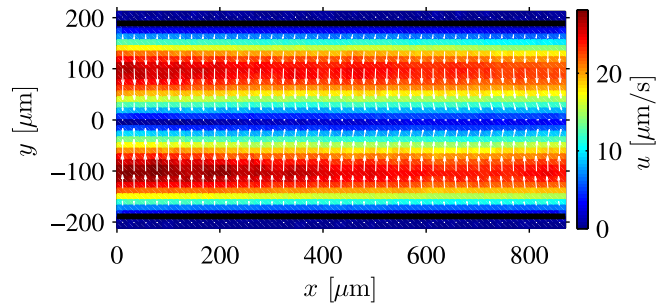}
    \caption{\figlab{upiv_example} Experimental micro-PIV measurement of acou\-sto\-pho\-resis in the horizontal ultrasound symmetry plane of 1-$\SImum$-diameter polystyrene particles suspended in water and driven at the piezo frequency $f=1.940~\SIMHz$ corresponding to $\lambda/2 = w$, temperature $25~\SIC$, and voltage $\Upp^*=7.94~\SIV$. The arrows represent the measured velocity vectors $\upvec$ and the colors their magnitude $\up$ normalized to $\Upp^*$, see \eqref{Upp_norm_upvec}.}
\end{figure}

\subsection{Excitation of a 1D transverse standing wave}
In \figsref{upiv_example}{vyavg_y_example} we verify experimentally that the acou\-sto\-pho\-retic particle velocity is of the predicted sinusoidal form given in \eqref{up} and resulting from a 1D transverse standing wave. For the actual applied voltage of $\Upp^*=7.94~\SIV$ the maximum velocity was measured to be $1.77~\SImmps$, which  by \eqref{Upp_norm_upvec} is normalized to the maximum velocity $\upstar_\mathrm{max}= 28~\SImumps$ seen in \figref{upiv_example}.

A detailed analysis of the measured velocity field reveals three main points: (i) The average of the ratio of the axial to the transverse velocity component is practically zero, $\langle |\up_x/\up_y| \rangle<5~\%$, (ii) the maximum particle velocity along any line with a given axial grid point coordinate $x_m$ varies less than $6~\%$ as a function of $x_m$, and (iii) the axial average $\langle\up_y\rangle_x$ of the transverse velocity component $\up_y$ is well fitted within small errorbars ($< 1$~\%) by \eqref{up}.

\begin{figure}[!t]
  \centering
  \includegraphics[width=85mm]{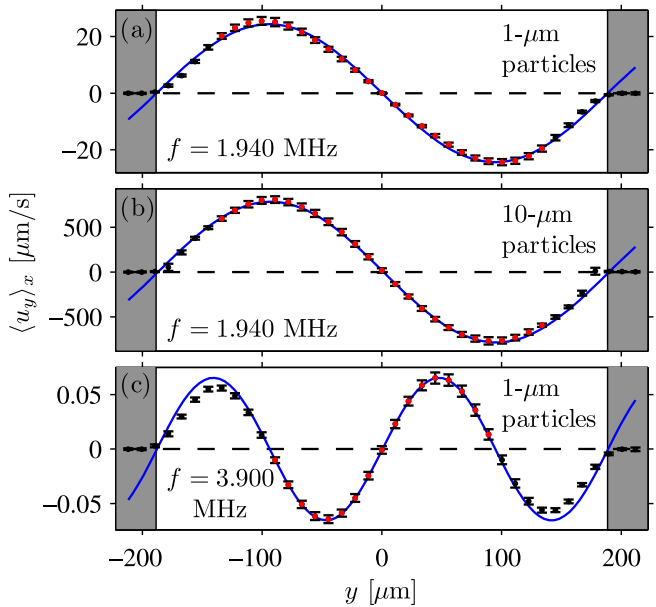}
\caption{\figlab{vyavg_y_example} Measured average  $\langle\up_y\rangle^{{}}_x$ (black and red points) and its standard deviation $\sigma(\langle\up_y\rangle^{{}}_x)$ (error bars) of the transverse velocity $\up_y$ of polystyrene particles in water. The sinusoidal prediction \eqref{up} for $\up$ (blue line) is fitted to data points $\langle\up_y\rangle^{{}}_x$ away from the side walls (red points). (a) Streaming-dominated acoustophoresis for $2a = 1~\SImum$ driven at $f=1.940~\SIMHz$ ($\lambda/2 = w$, same data as in \figref{upiv_example}). (b) Radiation-dominated acoustophoresis for $2a = 10~\SImum$ at $f=1.940~\SIMHz$ ($\lambda/2 = w$). (c) Streaming-dominated acoustophoresis  for $2a = 1~\SImum$ at $f=3.900~\SIMHz$ ($\lambda = w$).}
\end{figure}

\subsection{Measuring the velocity amplitude}
\seclab{meas_vel}

In \figref{vyavg_y_example}(a) we plot the axial average $\langle\up_y\rangle_x$ of the transverse velocity component $\up_y$ (black and red points) and its standard deviation $\sigma(\langle\up_y\rangle_x)$ (error bars) for the velocity field shown in \figref{upiv_example} at the standing half-wave resonance frequency $f=1.940~\SIMHz$ for the 1-$\SImum$-diameter streaming-dominated particles (series MQ1). The measured velocities away from the side walls (red points) are fitted well by the predicted sinusoidal velocity profile $\ua\sin(2ky)$ (blue curve) \eqref{up} for fixed wavelength $\lambda=2\pi/k=2w$ and using $\ua$ as the only fitting parameter. Velocities close to the side walls (black points) are discarded due to their interaction with the side walls. As seen numerically in \figref{ustr_along_y}, the no-slip boundary condition on the side walls of the rectangular geometry suppresses the streaming velocity near the side walls relative to sinusoidal velocity profile of the parallel-plate geometry.

As shown in \figref{vyavg_y_example}(b), the theoretical prediction also fits well the measured velocities away from the side walls for the large radiation-dominated 10-$\SImum$-diameter particles (series MQ1, $\lambda/2 = w$). Likewise, as seen in \figref{vyavg_y_example}(c), a good fit is also obtained for the  1-$\SImum$-diameter particles away from the side walls at the standing full-wave frequency $f=3.900~\SIMHz$ (series MQ2, $\lambda=w$).

Given this strong support for the presence of standing transverse waves, we use this standing-wave fitting procedure to determine the velocity amplitude $\ua$ in the following analysis of the acoustophoretic particle velocity.

In spite of the normalization to the same driving voltage of 1~V, the velocity amplitude of the half-wave resonance in \figref{vyavg_y_example}(a) is 400 times larger than that of the full-wave resonance in \figref{vyavg_y_example}(c). This is due to a difference in coupling to the piezo and in dissipation.

\begin{figure}[!t]
  \centering
    \includegraphics[width=85mm]{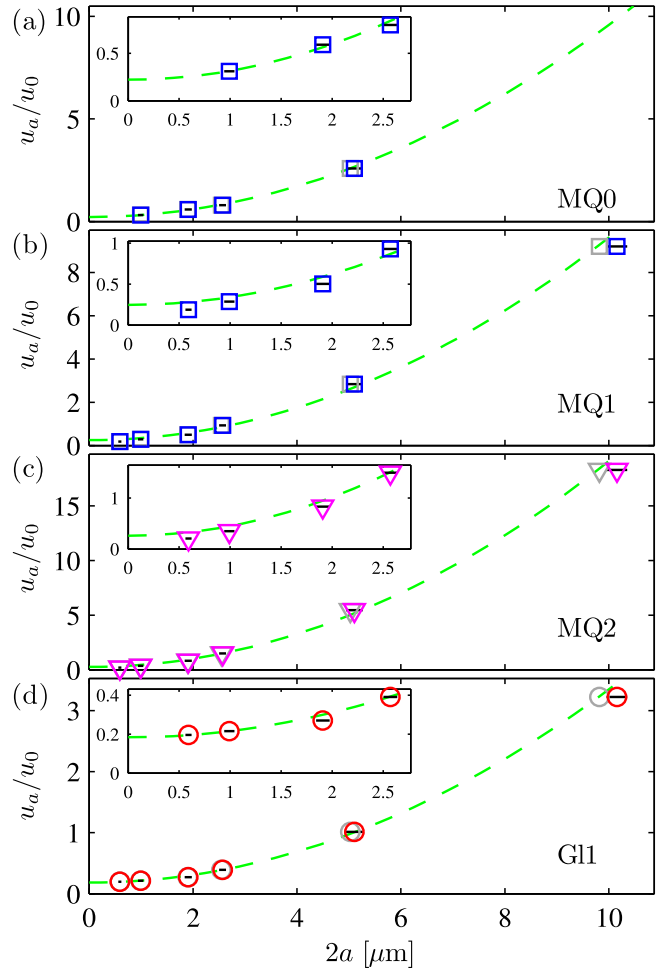}
	\caption{\figlab{uavsaall} Measured and normalized acoustophoretic velocity amplitudes $\ua/\uO$ as function of particle diameter $2a$ (colored symbols) and wall-drag-corrected particle diameter $2a_\mathrm{wd}$ (gray symbols) for the four experiment series (a) MQ0, (b) MQ1, (c) MQ2 and (d) Gl1 described in \secref{meas_series}. The characteristic velocity amplitude $\uO$ is determined from fitting \eqref{up} to the gray points in each series  using $\Eac$ and $s$ as fitting parameters resulting in the values listed in \tabref{fitresults}(a) (green dashed lines). The standard deviation of $\ua/\uO$ is less than the symbol size and the standard deviations on the particle diameters are indicated as black lines.\\[-5mm]}
\end{figure}

\begin{table}[!t]
\caption{\tablab{fitresults} Measured acoustic energy densities $\Eac$ normalized to $\Upp=1~\SIV$ and streaming coefficient $s$.}
\centering
\begin{ruledtabular}
\begin{tabular}{lcc}
\multicolumn{3}{l}{(a) Un-weighted fit to all points, see \figref{uavsaall}.} \\
Susp., freq. & $\Eac$ [\SIJpcm] & $s$\\\hline
MQ0, 1.940~MHz & 52.306 $\pm$ 0.918 & 0.222 $\pm$ 0.025 \\
MQ1, 1.940~MHz & 31.807 $\pm$ 0.569 & 0.247 $\pm$ 0.071 \\
MQ2, 3.900~MHz & 0.070 $\pm$ 0.001 & 0.262 $\pm$ 0.125 \\
Gl1, 2.027~MHz & 2.420 $\pm$ 0.020 & 0.184 $\pm$ 0.012 \\ \hline\hline
\multicolumn{3}{l}{(b) Based on particles with $2a=0.6~\SImum$ and $2a=10~\SImum$} \\
Susp., freq. & $\Eac$ [\SIJpcm]\footnote{\eqref{Eac}} & $s$\footnote{\eqref{s}}\\\hline
MQ1, 1.940~MHz & 32.436 $\pm$ 1.282 & 0.182 $\pm$ 0.008 \\
MQ2, 3.900~MHz & 0.071 $\pm$ 0.003 & 0.205 $\pm$ 0.008 \\
Gl1, 2.027~MHz & 2.559 $\pm$ 0.110 & 0.186 $\pm$ 0.008 \\
\end{tabular}
\end{ruledtabular}
\end{table}

\subsection{Velocity as function of particle diameter}
\seclab{velocity}

To analyze in detail the transverse velocity amplitude $\ua$ in all four series MQ0, MQ1, MQ2, and Gl1, we return to the wall-enhanced drag coefficient $\chi$ of \secref{wall_effects}. In general, $\chi$ depends in a non-linear way on the motion and position of the particle relative to the rigid walls. However, in \secref{microPIV_analysis} we established that the majority of the observed particles reside in the middle half of the channel, and in our standing-wave fitting procedure for $\ua$ in \secref{meas_vel} we discarded particles close to the side walls. Consequently, given this and the values of $\chi$ in \tabref{wall_corr}, it is a good approximation to assume that all involved particles have the same wall correction factor, namely the symmetry-plane, parallel-motion factor,
 \beq{chi_eff}
 \chi  \approx \chi^\mathrm{paral}_{z=0}.
 \eeq
As the drag-correction only enters on the radiation-induced term in \eqref{upvecwall}, we introduce a wall-drag-corrected particle size $a_\mathrm{wd}=(\chi^\mathrm{paral}_{z=0})^{-\frac{1}{2}}\: a$.

To determine the acoustic energy density $\Eac$ and the streaming coefficient $s$ we plot
in \figref{uavsaall}, for each of the four experiment series, $\ua/\uO$ versus the particle diameter $2a$ (colored symbols) and wall-drag-corrected particle diameter $2a_\mathrm{wd}$ (gray symbols). The characteristic velocity amplitude $\uO$ is determined in each series by fitting the wall-drag-corrected data points to \eqref{up} using $\Eac$ and $s$ as fitting parameters. In all four experiment series, a clear $a^2$-dependence is seen. Notice further that the velocities follow almost the same distribution around the fitted line in all series. This we suspect may be due to systematic errors, \eg\  that the 5-$\SImum$-diameter particles are slightly underestimated (see the Coulter data in Supplemental Material). The resulting fitting parameters $\Eac$ and $s$ are listed in \tabref{fitresults}(a). The energy densities normalized to $\Upp=1~\SIV$, see \eqref{Eac_norm} varies with more than a factor 700 due to a large difference in the strength of the excited resonances. According to the predictions in \secref{SingleParticle} the streaming coefficient $s$ should be constant, but experimentally it varies from $0.18$ to $0.25$. However, taking the fitting uncertainties into account in a weighted average, leads to $\langle s\rangle_\mathrm{w}=0.192\pm 0.010$ close to $s^T_\mathrm{r} \approx 0.194$ of \eqref{ustr1DT}.

Another approach for extracting $\Eac$ and $s$ is to assume that the smallest particles $2 a = 0.6~\SImum$ are influenced only by the streaming-induced drag. If so, the velocity of the largest $2 a = 10~\SImum$ particle has a streaming component of less than $6~\%$, see the measured ratios $\ua^{0.6~\SImum}/\ua^{10~\SImum}$ in \tabref{velresults}. Therefore, we further assume that the $10~\SImum$-diameter particles are influenced solely by the radiation force, and from $\uarad=\ua(a/a_0)^2$ we determine the acoustic energy density as
\bal
\Eac  = \frac{3}{2}\frac{\eta\cO}{\Phi\omega}\frac{\ua^{10~\SImum}}{a_\mathrm{wd}^2}\eqlab{Eac}.
\eal
Knowing the acoustic energy density, we use \eqref{ustr1D} to calculate the streaming coefficient $s$ from $\uastr=\uO s$ as
\bal
	s &= \frac{\rhoO\cO}{4\Eac}u^\mathrm{0.6~\SImum}_a\eqlab{s}.
\eal
Assuming that the largest error is due to the dispersion in particle size, we obtain the results listed in \tabref{fitresults}(b). The acoustic energy densities are close to the ones extracted from the fits in \figref{uavsaall} and the geometric streaming coefficient varies from 0.180 to 0.203 with an weighted average of $\langle s\rangle_\mathrm{w}=0.191\pm 0.005$. Note that using \eqsref{Eac}{s}, we only need to consider the dispersion of the 10-$\SImum$-diameter particles, which results in a more reliable estimate of $s$.

\begin{figure}[!t]
  \centering
    \includegraphics[width=85mm]{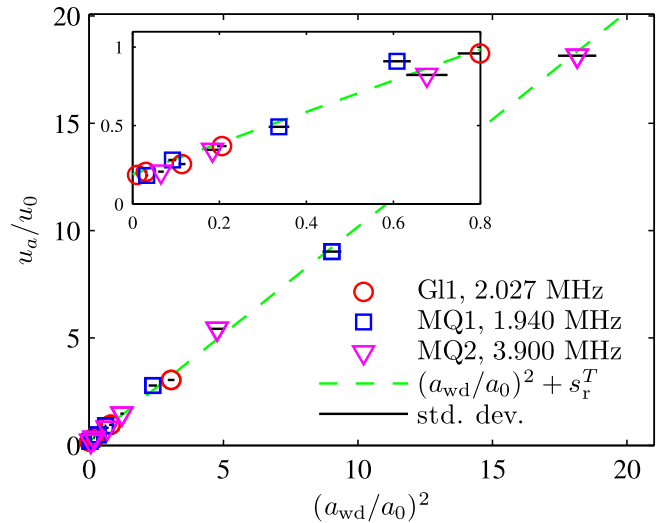}
	\caption{\figlab{uacollapse} Normalized acoustophoretic particle velocities $\ua/\uO$ versus normalized particle size squared $(a_\mathrm{wd}/a_0)^2$.}
\end{figure}

We use the acoustic energy densities in \tabref{fitresults}(b) together with the material parameters in \tabref{parameters} to calculate $\uO$ and $a_0$, \eqref{u0_a0}, for each of the experiment series MQ1, MQ2, and Gl1. According to the theoretical prediction in \eqref{up}, all data points must fall on a straight line of unity slope and intersection $s$ if plotted as the normalized velocity amplitude $\ua/\uO$ as function of the normalized particle radius squared $(a/a_0)^2$. The plot is shown in \figref{uacollapse} showing good agreement with the theoretical prediction using $s^T_\mathrm{r} \approx 0.194$.

\begin{table}[!t]
\caption{\label{tab:velresults} Relative particle velocities.}
\centering
\begin{ruledtabular}
\begin{tabular}{lccc}
Susp., freq. & $\frac{\ua^{0.6~\SImum}}{\ua^{10~\SImum}}$ & $\frac{\ua^{10~\SImum}}{\ua^{0.6~\SImum}}$ & $\frac{\ua^{5~\SImum}-\ua^{0.6~\SImum}}{\ua^{10~\SImum}-\ua^{0.6~\SImum}}$ \\\hline
MQ1, 1.940~MHz & 0.020 &     49.6 & 0.294 \\
MQ2, 3.900~MHz & 0.011 &       88.4 & 0.291 \\
Gl1, 2.027~MHz & 0.061 &         16.4 & 0.270 \\
\end{tabular}
\end{ruledtabular}
\end{table}

\subsection{Velocity ratios}

In \tabref{velresults} we list velocity ratios for different particle sizes in the experiment series MQ1, MQ2, and Gl1.

From  \eqref{up} we expect $\ua-\uastr\propto a^2$ leading to the prediction $(\ua^{5~\SImum}-\uastr)/(\ua^{10~\SImum}-\uastr)=(5/10)^2=0.25$. If we assume that the smallest 0.6-$\SImum$-diameter particles are only influenced by the acoustic streaming, we have $\uastr=\ua^{0.6~\SImum}$. We can therefore test the just mentioned hypothesis by calculating $(\ua^{5~\SImum}-\ua^{0.6~\SImum})/(\ua^{10~\SImum}-\ua^{0.6~\SImum})$. The results are listed in the third column in \tabref{velresults}, where we obtain values ranging from 0.27 to to 0.29, or a deviation of 8 to 18~\%.

Assuming that the smallest 0.6-$\SImum$-diameter particles and the largest 10-$\SImum$-diameter particles are influenced only by the acoustic streaming and the acoustic radiation force, respectively, we can estimate the ratio of radiation- and streaming-induced velocities as $\uarad/\uastr=\ua^{10~\SImum}/\ua^{0.6~\SImum}$, which are listed in the second column in \tabref{velresults}. First, we notice that the ratio increases by a factor of $88.4/49.6=1.8$ as we increase the frequency by a factor of $3.900/1.940=2.0$. This agrees well with a linear increase with frequency as predicted by \eqref{ratio}. Secondly, we notice that the ratio increases by a factor of $48.6/16.4=3.0$ as we change the suspending medium from a 0.75:0.25 mixture of water and glycerol to pure water. According to \eqref{ratio} $\uarad/\uastr$ increases linearly with $\nu/\Phi$ and from \tabref{parameters} we obtain a predicted ratio increase of $16.8/5.25=3.2$, which matches well with the experimentally-estimated ratio.

With these results we have gained experimental support for the theoretical prediction of the velocity ratio given in \eqref{ratio}.

\section{Discussion}
\seclab{discussion}

Our results verify experimentally the theoretically predicted dependence of the magnitude of the acoustophoretic velocity in a microchannel on the viscosity of the suspending liquid, the acoustic contrast factor $\Phi$, and the ultrasound frequency.

For most situations involving cells, isotonic solutions are used such as PBS, sodium chloride, or blood plasma. Direct manipulation of small particles such as bacteria in plasma or in buffers of high levels of protein is problematic, primarily due to the high viscosity of those media. When possible, these media should be exchanged prior to manipulation to increase the potential for success.

Increasing the frequency in the system would allow for a reduction of the critical diameter of particles. One adverse effect of a higher frequency is that the channel width must be narrowed down which affect the throughput in the system. This problem can, however, be overcome by designing a channel of high aspect ratio, where the resonance occurs over the smallest dimension \cite{Adams2012}.

Another benefit of high aspect ratio channels was pointed out by Muller \etal\ \cite{Muller2012}. Since the acoustic streaming emanates from the walls perpendicular to the wave propagation, here the top and bottom, a high channel leads to a weaker average streaming field in the center.

The measurements of particle velocities for polystyrene particles ranging from $0.6~\SImum$ to $10~\SImum$ give no support to previous measurements presented by Yasuda and Kamakura \cite{Yasuda1997} in 1997. Their rather spectacular result was that particles below a certain size move faster than larger particles do. From the experiments reported herein it is clear, however, that the motion of particles indeed can be well described with the analyses presented by Rayleigh \cite{LordRayleigh1884}, Yosioka and Kawasima  \cite{Yosioka1955}, and Gorkov \cite{Gorkov1962}.

The uncertainties in the measured particle velocities may in particular be due to the following four causes: (\textit{i}) Variations in particle density $\rhoP$ and compressibility $\KP$ as function of particle producer (Fluka and G. Kisker) or batch, (\textit{ii}) deviations from normal distributed  particle sizes as shown in Supplemental Material Fig.~1, (\textit{ii}) local fluctuations in the bead concentrations leading to particle-particle interactions, and (\textit{iv}) viscosity variations induced by the suspended particles.

To better understand the nature of acoustic streaming in microchannels the streaming field should be mapped for the channel cross section and along the whole length of the channel. As reported by Hagsäter \etal\  \cite{Hagsater2008} and Augustsson \etal\ \cite{Augustsson2011} the acoustic field can deviate dramatically from the very uniform one dimensional field reported herein. Non symmetrical acoustic fields can be expected to generate far more complex streaming fields.

\section{Conclusions}
\seclab{conclusions}

We have investigated the motion of microparticles due to acoustic radiation and acoustic streaming inside a liquid-filled long, straight rectangular channel of width $w$ and height $h$ driven by an ultrasound standing wave of wavelength $\lambda$ .

Fortuitously, the simple analytical expression derived by Lord Rayleigh for the streaming velocity in an isothermal liquid slab between two infinite parallel plates fulfilling $\lambda \gg h$, is a good approximation for the specific rectangular channel of \figref{chip_sketch} containing a thermoviscous liquid and fulfilling $\lambda \approx h$. The reduction in velocity obtained when substituting the parallel plates with the rectangular geometry is almost perfectly compensated for by the enhancement in velocity from substituting the isothermal liquid by the thermoviscous one.

A theoretical prediction was made, \eqref{up}, for the dependence of the radiation- and streaming-induced velocities on the size of the particles, the ultrasound frequency, the viscosity of the suspending liquid, and the acoustic contrast factor. This prediction was found to be in excellent agreement with experimental findings as shown by the collapse after re-scaling of data from 22 different measurement on the same line in \figref{uacollapse}. The results have bearing on acoustophoretic manipulation strategies for sub-micrometer biological particles such as bacteria and vira, which are too small to be handled using the present manifestation of this technique. We can conclude that increasing the ultrasound frequency, increase of the channel aspect ratio, and lowering the viscosity of the suspending fluid is probably the most viable route to conduct such manipulation.

\section*{Acknowledgements}
This research was supported by the Danish Council for Independent
Research, Technology and Production Sciences, Grant No.~274-09-0342;
the Swedish Research Council, Grant No.~2007-4946; and the Swedish
Governmental Agency for Innovation Systems, VINNOVA, the program
Innovations for Future Health, Cell CARE, Grant No.~2009-00236.


\begin{thebibliography}{41}%
\makeatletter
\providecommand \@ifxundefined [1]{%
 \@ifx{#1\undefined}
}%
\providecommand \@ifnum [1]{%
 \ifnum #1\expandafter \@firstoftwo
 \else \expandafter \@secondoftwo
 \fi
}%
\providecommand \@ifx [1]{%
 \ifx #1\expandafter \@firstoftwo
 \else \expandafter \@secondoftwo
 \fi
}%
\providecommand \natexlab [1]{#1}%
\providecommand \enquote  [1]{``#1''}%
\providecommand \bibnamefont  [1]{#1}%
\providecommand \bibfnamefont [1]{#1}%
\providecommand \citenamefont [1]{#1}%
\providecommand \href@noop [0]{\@secondoftwo}%
\providecommand \href [0]{\begingroup \@sanitize@url \@href}%
\providecommand \@href[1]{\@@startlink{#1}\@@href}%
\providecommand \@@href[1]{\endgroup#1\@@endlink}%
\providecommand \@sanitize@url [0]{\catcode `\\12\catcode `\$12\catcode
  `\&12\catcode `\#12\catcode `\^12\catcode `\_12\catcode `\%12\relax}%
\providecommand \@@startlink[1]{}%
\providecommand \@@endlink[0]{}%
\providecommand \url  [0]{\begingroup\@sanitize@url \@url }%
\providecommand \@url [1]{\endgroup\@href {#1}{\urlprefix }}%
\providecommand \urlprefix  [0]{URL }%
\providecommand \Eprint [0]{\href }%
\providecommand \doibase [0]{http://dx.doi.org/}%
\providecommand \selectlanguage [0]{\@gobble}%
\providecommand \bibinfo  [0]{\@secondoftwo}%
\providecommand \bibfield  [0]{\@secondoftwo}%
\providecommand \translation [1]{[#1]}%
\providecommand \BibitemOpen [0]{}%
\providecommand \bibitemStop [0]{}%
\providecommand \bibitemNoStop [0]{.\EOS\space}%
\providecommand \EOS [0]{\spacefactor3000\relax}%
\providecommand \BibitemShut  [1]{\csname bibitem#1\endcsname}%
\let\auto@bib@innerbib\@empty
\bibitem [{\citenamefont {Friend}\ and\ \citenamefont
  {Yeo}(2011)}]{Friend2011}%
  \BibitemOpen
  \bibfield  {author} {\bibinfo {author} {\bibfnamefont {J.}~\bibnamefont
  {Friend}}\ and\ \bibinfo {author} {\bibfnamefont {L.~Y.}\ \bibnamefont
  {Yeo}},\ }\href {\doibase 10.1103/RevModPhys.83.647} {\bibfield  {journal}
  {\bibinfo  {journal} {Rev Mod Phys}\ }\textbf {\bibinfo {volume} {83}},\
  \bibinfo {pages} {647} (\bibinfo {year} {2011})}\BibitemShut {NoStop}%
\bibitem [{\citenamefont {Bruus}\ \emph {et~al.}(2011)\citenamefont {Bruus},
  \citenamefont {Dual}, \citenamefont {Hawkes}, \citenamefont {Hill},
  \citenamefont {Laurell}, \citenamefont {Nilsson}, \citenamefont {Radel},
  \citenamefont {Sadhal},\ and\ \citenamefont {Wiklund}}]{Bruus2011c}%
  \BibitemOpen
  \bibfield  {author} {\bibinfo {author} {\bibfnamefont {H.}~\bibnamefont
  {Bruus}}, \bibinfo {author} {\bibfnamefont {J.}~\bibnamefont {Dual}},
  \bibinfo {author} {\bibfnamefont {J.}~\bibnamefont {Hawkes}}, \bibinfo
  {author} {\bibfnamefont {M.}~\bibnamefont {Hill}}, \bibinfo {author}
  {\bibfnamefont {T.}~\bibnamefont {Laurell}}, \bibinfo {author} {\bibfnamefont
  {J.}~\bibnamefont {Nilsson}}, \bibinfo {author} {\bibfnamefont
  {S.}~\bibnamefont {Radel}}, \bibinfo {author} {\bibfnamefont
  {S.}~\bibnamefont {Sadhal}}, \ and\ \bibinfo {author} {\bibfnamefont
  {M.}~\bibnamefont {Wiklund}},\ }\href {\doibase 10.1039/c1lc90058g}
  {\bibfield  {journal} {\bibinfo  {journal} {Lab Chip}\ }\textbf {\bibinfo
  {volume} {11}},\ \bibinfo {pages} {3579} (\bibinfo {year}
  {2011})}\BibitemShut {NoStop}%
\bibitem [{\citenamefont {Hawkes}\ \emph {et~al.}(2004)\citenamefont {Hawkes},
  \citenamefont {Barber}, \citenamefont {Emerson},\ and\ \citenamefont
  {Coakley}}]{Hawkes2004}%
  \BibitemOpen
  \bibfield  {author} {\bibinfo {author} {\bibfnamefont {J.~J.}\ \bibnamefont
  {Hawkes}}, \bibinfo {author} {\bibfnamefont {R.~W.}\ \bibnamefont {Barber}},
  \bibinfo {author} {\bibfnamefont {D.~R.}\ \bibnamefont {Emerson}}, \ and\
  \bibinfo {author} {\bibfnamefont {W.~T.}\ \bibnamefont {Coakley}},\ }\href
  {\doibase 10.1039/B408045A} {\bibfield  {journal} {\bibinfo  {journal} {Lab
  Chip}\ }\textbf {\bibinfo {volume} {4}},\ \bibinfo {pages} {446} (\bibinfo
  {year} {2004})}\BibitemShut {NoStop}%
\bibitem [{\citenamefont {Petersson}\ \emph {et~al.}(2004)\citenamefont
  {Petersson}, \citenamefont {Nilsson}, \citenamefont {Holm}, \citenamefont
  {J\"{o}nsson},\ and\ \citenamefont {Laurell}}]{Petersson2004}%
  \BibitemOpen
  \bibfield  {author} {\bibinfo {author} {\bibfnamefont {F.}~\bibnamefont
  {Petersson}}, \bibinfo {author} {\bibfnamefont {A.}~\bibnamefont {Nilsson}},
  \bibinfo {author} {\bibfnamefont {C.}~\bibnamefont {Holm}}, \bibinfo {author}
  {\bibfnamefont {H.}~\bibnamefont {J\"{o}nsson}}, \ and\ \bibinfo {author}
  {\bibfnamefont {T.}~\bibnamefont {Laurell}},\ }\href {\doibase
  10.1039/b409139f} {\bibfield  {journal} {\bibinfo  {journal} {Analyst}\
  }\textbf {\bibinfo {volume} {129}},\ \bibinfo {pages} {938} (\bibinfo {year}
  {2004})}\BibitemShut {NoStop}%
\bibitem [{\citenamefont {Thevoz}\ \emph {et~al.}(2010)\citenamefont {Thevoz},
  \citenamefont {Adams}, \citenamefont {Shea}, \citenamefont {Bruus},\ and\
  \citenamefont {Soh}}]{Thevoz2010}%
  \BibitemOpen
  \bibfield  {author} {\bibinfo {author} {\bibfnamefont {P.}~\bibnamefont
  {Thevoz}}, \bibinfo {author} {\bibfnamefont {J.~D.}\ \bibnamefont {Adams}},
  \bibinfo {author} {\bibfnamefont {H.}~\bibnamefont {Shea}}, \bibinfo {author}
  {\bibfnamefont {H.}~\bibnamefont {Bruus}}, \ and\ \bibinfo {author}
  {\bibfnamefont {H.~T.}\ \bibnamefont {Soh}},\ }\href {\doibase
  10.1021/ac100357u} {\bibfield  {journal} {\bibinfo  {journal} {Anal Chem}\
  }\textbf {\bibinfo {volume} {82}},\ \bibinfo {pages} {3094} (\bibinfo {year}
  {2010})}\BibitemShut {NoStop}%
\bibitem [{\citenamefont {Augustsson}\ \emph {et~al.}(2012)\citenamefont
  {Augustsson}, \citenamefont {Magnusson}, \citenamefont {Nordin},
  \citenamefont {Lilja},\ and\ \citenamefont {Laurell}}]{Augustsson2012b}%
  \BibitemOpen
  \bibfield  {author} {\bibinfo {author} {\bibfnamefont {P.}~\bibnamefont
  {Augustsson}}, \bibinfo {author} {\bibfnamefont {C.}~\bibnamefont
  {Magnusson}}, \bibinfo {author} {\bibfnamefont {M.}~\bibnamefont {Nordin}},
  \bibinfo {author} {\bibfnamefont {H.}~\bibnamefont {Lilja}}, \ and\ \bibinfo
  {author} {\bibfnamefont {T.}~\bibnamefont {Laurell}},\ }\href@noop {}
  {\bibfield  {journal} {\bibinfo  {journal} {Anal. Chem.}\ }\textbf {\bibinfo
  {volume} {in press}} (\bibinfo {year} {2012})}\BibitemShut {NoStop}%
\bibitem [{\citenamefont {Ding}\ \emph {et~al.}(2012)\citenamefont {Ding},
  \citenamefont {Lin}, \citenamefont {Kiraly}, \citenamefont {Yue},
  \citenamefont {Li}, \citenamefont {Chiang}, \citenamefont {Shi},
  \citenamefont {Benkovic},\ and\ \citenamefont {Huang}}]{Ding2012}%
  \BibitemOpen
  \bibfield  {author} {\bibinfo {author} {\bibfnamefont {X.}~\bibnamefont
  {Ding}}, \bibinfo {author} {\bibfnamefont {S.-C.~S.}\ \bibnamefont {Lin}},
  \bibinfo {author} {\bibfnamefont {B.}~\bibnamefont {Kiraly}}, \bibinfo
  {author} {\bibfnamefont {H.}~\bibnamefont {Yue}}, \bibinfo {author}
  {\bibfnamefont {S.}~\bibnamefont {Li}}, \bibinfo {author} {\bibfnamefont
  {I.-K.}\ \bibnamefont {Chiang}}, \bibinfo {author} {\bibfnamefont
  {J.}~\bibnamefont {Shi}}, \bibinfo {author} {\bibfnamefont {S.~J.}\
  \bibnamefont {Benkovic}}, \ and\ \bibinfo {author} {\bibfnamefont {T.~J.}\
  \bibnamefont {Huang}},\ }\href {\doibase 10.1073/pnas.1209288109} {\bibfield
  {journal} {\bibinfo  {journal} {PNAS}\ }\textbf {\bibinfo {volume} {109}},\
  \bibinfo {pages} {11105} (\bibinfo {year} {2012})}\BibitemShut {NoStop}%
\bibitem [{\citenamefont {Vanherberghen}\ \emph {et~al.}(2010)\citenamefont
  {Vanherberghen}, \citenamefont {Manneberg}, \citenamefont {Christakou},
  \citenamefont {Frisk}, \citenamefont {Ohlin}, \citenamefont {Hertz},
  \citenamefont {Onfelt},\ and\ \citenamefont {Wiklund}}]{Vanherberghen2010}%
  \BibitemOpen
  \bibfield  {author} {\bibinfo {author} {\bibfnamefont {B.}~\bibnamefont
  {Vanherberghen}}, \bibinfo {author} {\bibfnamefont {O.}~\bibnamefont
  {Manneberg}}, \bibinfo {author} {\bibfnamefont {A.}~\bibnamefont
  {Christakou}}, \bibinfo {author} {\bibfnamefont {T.}~\bibnamefont {Frisk}},
  \bibinfo {author} {\bibfnamefont {M.}~\bibnamefont {Ohlin}}, \bibinfo
  {author} {\bibfnamefont {H.~M.}\ \bibnamefont {Hertz}}, \bibinfo {author}
  {\bibfnamefont {B.}~\bibnamefont {Onfelt}}, \ and\ \bibinfo {author}
  {\bibfnamefont {M.}~\bibnamefont {Wiklund}},\ }\href {\doibase
  10.1039/c004707d} {\bibfield  {journal} {\bibinfo  {journal} {Lab Chip}\
  }\textbf {\bibinfo {volume} {10}},\ \bibinfo {pages} {2727} (\bibinfo {year}
  {2010})}\BibitemShut {NoStop}%
\bibitem [{\citenamefont {Augustsson}\ and\ \citenamefont
  {Laurell}(2012)}]{Augustsson2012a}%
  \BibitemOpen
  \bibfield  {author} {\bibinfo {author} {\bibfnamefont {P.}~\bibnamefont
  {Augustsson}}\ and\ \bibinfo {author} {\bibfnamefont {T.}~\bibnamefont
  {Laurell}},\ }\href {\doibase 10.1039/c2lc40200a} {\bibfield  {journal}
  {\bibinfo  {journal} {Lab Chip}\ }\textbf {\bibinfo {volume} {12}},\ \bibinfo
  {pages} {1742} (\bibinfo {year} {2012})}\BibitemShut {NoStop}%
\bibitem [{\citenamefont {Bengtsson}\ and\ \citenamefont
  {Laurell}(2004)}]{Bengtsson2004}%
  \BibitemOpen
  \bibfield  {author} {\bibinfo {author} {\bibfnamefont {M.}~\bibnamefont
  {Bengtsson}}\ and\ \bibinfo {author} {\bibfnamefont {T.}~\bibnamefont
  {Laurell}},\ }\href {\doibase 10.1007/s00216-003-2334-y} {\bibfield
  {journal} {\bibinfo  {journal} {Anal Bioanal Chem}\ }\textbf {\bibinfo
  {volume} {378}},\ \bibinfo {pages} {1716} (\bibinfo {year}
  {2004})}\BibitemShut {NoStop}%
\bibitem [{\citenamefont {Kuznetsova}\ \emph {et~al.}(2005)\citenamefont
  {Kuznetsova}, \citenamefont {Martin},\ and\ \citenamefont
  {Coakley}}]{Kuznetsova2005}%
  \BibitemOpen
  \bibfield  {author} {\bibinfo {author} {\bibfnamefont {L.~a.}\ \bibnamefont
  {Kuznetsova}}, \bibinfo {author} {\bibfnamefont {S.~P.}\ \bibnamefont
  {Martin}}, \ and\ \bibinfo {author} {\bibfnamefont {W.~T.}\ \bibnamefont
  {Coakley}},\ }\href {\doibase 10.1016/j.bios.2005.02.014} {\bibfield
  {journal} {\bibinfo  {journal} {Biosensors \& bioelectronics}\ }\textbf
  {\bibinfo {volume} {21}},\ \bibinfo {pages} {940} (\bibinfo {year}
  {2005})}\BibitemShut {NoStop}%
\bibitem [{\citenamefont {Martin}\ \emph {et~al.}(2005)\citenamefont {Martin},
  \citenamefont {Towsend}, \citenamefont {Kuznetsova}, \citenamefont
  {Borthwick}, \citenamefont {Hill}, \citenamefont {McDonnell},\ and\
  \citenamefont {Coakley}}]{Martin2005}%
  \BibitemOpen
  \bibfield  {author} {\bibinfo {author} {\bibfnamefont {S.~P.}\ \bibnamefont
  {Martin}}, \bibinfo {author} {\bibfnamefont {R.~J.}\ \bibnamefont {Towsend}},
  \bibinfo {author} {\bibfnamefont {L.~A.}\ \bibnamefont {Kuznetsova}},
  \bibinfo {author} {\bibfnamefont {K.~A.~J.}\ \bibnamefont {Borthwick}},
  \bibinfo {author} {\bibfnamefont {M.}~\bibnamefont {Hill}}, \bibinfo {author}
  {\bibfnamefont {M.~B.}\ \bibnamefont {McDonnell}}, \ and\ \bibinfo {author}
  {\bibfnamefont {W.~T.}\ \bibnamefont {Coakley}},\ }\href@noop {} {\bibfield
  {journal} {\bibinfo  {journal} {Biosens Bioelectron}\ }\textbf {\bibinfo
  {volume} {21}},\ \bibinfo {pages} {758} (\bibinfo {year} {2005})}\BibitemShut
  {NoStop}%
\bibitem [{\citenamefont {Hammarstrom}\ \emph {et~al.}(697G)\citenamefont
  {Hammarstrom}, \citenamefont {Laurell},\ and\ \citenamefont
  {Nilsson}}]{Hammarstrom2012}%
  \BibitemOpen
  \bibfield  {author} {\bibinfo {author} {\bibfnamefont {B.}~\bibnamefont
  {Hammarstrom}}, \bibinfo {author} {\bibfnamefont {T.}~\bibnamefont
  {Laurell}}, \ and\ \bibinfo {author} {\bibfnamefont {J.}~\bibnamefont
  {Nilsson}},\ }\href {\doibase 10.1039/C2LC40697G} {\bibfield  {journal}
  {\bibinfo  {journal} {Lab Chip}\ }\textbf {\bibinfo {volume} {12}},\ \bibinfo
  {pages} {in press} (\bibinfo {year} {2012, doi:
  10.1039/C2LC40697G})}\BibitemShut {NoStop}%
\bibitem [{\citenamefont {Barnkob}\ \emph {et~al.}(2010)\citenamefont
  {Barnkob}, \citenamefont {Augustsson}, \citenamefont {Laurell},\ and\
  \citenamefont {Bruus}}]{Barnkob2010}%
  \BibitemOpen
  \bibfield  {author} {\bibinfo {author} {\bibfnamefont {R.}~\bibnamefont
  {Barnkob}}, \bibinfo {author} {\bibfnamefont {P.}~\bibnamefont {Augustsson}},
  \bibinfo {author} {\bibfnamefont {T.}~\bibnamefont {Laurell}}, \ and\
  \bibinfo {author} {\bibfnamefont {H.}~\bibnamefont {Bruus}},\ }\href
  {\doibase 10.1039/b920376a} {\bibfield  {journal} {\bibinfo  {journal} {Lab
  Chip}\ }\textbf {\bibinfo {volume} {10}},\ \bibinfo {pages} {563} (\bibinfo
  {year} {2010})}\BibitemShut {NoStop}%
\bibitem [{\citenamefont {Augustsson}\ \emph {et~al.}(2011)\citenamefont
  {Augustsson}, \citenamefont {Barnkob}, \citenamefont {Wereley}, \citenamefont
  {Bruus},\ and\ \citenamefont {Laurell}}]{Augustsson2011}%
  \BibitemOpen
  \bibfield  {author} {\bibinfo {author} {\bibfnamefont {P.}~\bibnamefont
  {Augustsson}}, \bibinfo {author} {\bibfnamefont {R.}~\bibnamefont {Barnkob}},
  \bibinfo {author} {\bibfnamefont {S.~T.}\ \bibnamefont {Wereley}}, \bibinfo
  {author} {\bibfnamefont {H.}~\bibnamefont {Bruus}}, \ and\ \bibinfo {author}
  {\bibfnamefont {T.}~\bibnamefont {Laurell}},\ }\href {\doibase
  10.1039/c1lc20637k} {\bibfield  {journal} {\bibinfo  {journal} {Lab Chip}\
  }\textbf {\bibinfo {volume} {11}},\ \bibinfo {pages} {4152} (\bibinfo {year}
  {2011})}\BibitemShut {NoStop}%
\bibitem [{\citenamefont {Barnkob}\ \emph {et~al.}(2012)\citenamefont
  {Barnkob}, \citenamefont {Iranmanesh}, \citenamefont {Wiklund},\ and\
  \citenamefont {Bruus}}]{Barnkob2012}%
  \BibitemOpen
  \bibfield  {author} {\bibinfo {author} {\bibfnamefont {R.}~\bibnamefont
  {Barnkob}}, \bibinfo {author} {\bibfnamefont {I.}~\bibnamefont {Iranmanesh}},
  \bibinfo {author} {\bibfnamefont {M.}~\bibnamefont {Wiklund}}, \ and\
  \bibinfo {author} {\bibfnamefont {H.}~\bibnamefont {Bruus}},\ }\href
  {\doibase 10.1039/C2LC40120G} {\bibfield  {journal} {\bibinfo  {journal} {Lab
  Chip}\ }\textbf {\bibinfo {volume} {12}},\ \bibinfo {pages} {2337} (\bibinfo
  {year} {2012})}\BibitemShut {NoStop}%
\bibitem [{\citenamefont {Yosioka}\ and\ \citenamefont
  {Kawasima}(1955)}]{Yosioka1955}%
  \BibitemOpen
  \bibfield  {author} {\bibinfo {author} {\bibfnamefont {K.}~\bibnamefont
  {Yosioka}}\ and\ \bibinfo {author} {\bibfnamefont {Y.}~\bibnamefont
  {Kawasima}},\ }\href@noop {} {\bibfield  {journal} {\bibinfo  {journal}
  {Acustica}\ }\textbf {\bibinfo {volume} {5}},\ \bibinfo {pages} {167}
  (\bibinfo {year} {1955})}\BibitemShut {NoStop}%
\bibitem [{\citenamefont {Gorkov}(1962)}]{Gorkov1962}%
  \BibitemOpen
  \bibfield  {author} {\bibinfo {author} {\bibfnamefont {L.~P.}\ \bibnamefont
  {Gorkov}},\ }\href@noop {} {\bibfield  {journal} {\bibinfo  {journal} {Soviet
  Physics - Doklady}\ }\textbf {\bibinfo {volume} {6}},\ \bibinfo {pages} {773}
  (\bibinfo {year} {1962})}\BibitemShut {NoStop}%
\bibitem [{\citenamefont {Wiklund}\ \emph {et~al.}(2012)\citenamefont
  {Wiklund}, \citenamefont {Green},\ and\ \citenamefont {Ohlin}}]{Wiklund2012}%
  \BibitemOpen
  \bibfield  {author} {\bibinfo {author} {\bibfnamefont {M.}~\bibnamefont
  {Wiklund}}, \bibinfo {author} {\bibfnamefont {R.}~\bibnamefont {Green}}, \
  and\ \bibinfo {author} {\bibfnamefont {M.}~\bibnamefont {Ohlin}},\
  }\href@noop {} {\bibfield  {journal} {\bibinfo  {journal} {Lab Chip}\
  }\textbf {\bibinfo {volume} {12}},\ \bibinfo {pages} {2438} (\bibinfo {year}
  {2012})}\BibitemShut {NoStop}%
\bibitem [{\citenamefont {Spengler}\ \emph {et~al.}(2003)\citenamefont
  {Spengler}, \citenamefont {Coakley},\ and\ \citenamefont
  {Christensen}}]{spengler2003}%
  \BibitemOpen
  \bibfield  {author} {\bibinfo {author} {\bibfnamefont {J.~F.}\ \bibnamefont
  {Spengler}}, \bibinfo {author} {\bibfnamefont {W.~T.}\ \bibnamefont
  {Coakley}}, \ and\ \bibinfo {author} {\bibfnamefont {K.~T.}\ \bibnamefont
  {Christensen}},\ }\href {\doibase 10.1002/aic.690491110} {\bibfield
  {journal} {\bibinfo  {journal} {AIChE J}\ }\textbf {\bibinfo {volume} {49}},\
  \bibinfo {pages} {2773} (\bibinfo {year} {2003})}\BibitemShut {NoStop}%
\bibitem [{\citenamefont {Hags\"{a}ter}\ \emph {et~al.}(2007)\citenamefont
  {Hags\"{a}ter}, \citenamefont {Jensen}, \citenamefont {Bruus},\ and\
  \citenamefont {Kutter}}]{Hagsater2007}%
  \BibitemOpen
  \bibfield  {author} {\bibinfo {author} {\bibfnamefont {S.~M.}\ \bibnamefont
  {Hags\"{a}ter}}, \bibinfo {author} {\bibfnamefont {T.~G.}\ \bibnamefont
  {Jensen}}, \bibinfo {author} {\bibfnamefont {H.}~\bibnamefont {Bruus}}, \
  and\ \bibinfo {author} {\bibfnamefont {J.~P.}\ \bibnamefont {Kutter}},\
  }\href {\doibase 10.1039/b704864e} {\bibfield  {journal} {\bibinfo  {journal}
  {Lab Chip}\ }\textbf {\bibinfo {volume} {7}},\ \bibinfo {pages} {1336}
  (\bibinfo {year} {2007})}\BibitemShut {NoStop}%
\bibitem [{\citenamefont {Settnes}\ and\ \citenamefont
  {Bruus}(2012)}]{Settnes2012}%
  \BibitemOpen
  \bibfield  {author} {\bibinfo {author} {\bibfnamefont {M.}~\bibnamefont
  {Settnes}}\ and\ \bibinfo {author} {\bibfnamefont {H.}~\bibnamefont
  {Bruus}},\ }\href {\doibase 10.1103/PhysRevE.85.016327} {\bibfield  {journal}
  {\bibinfo  {journal} {Phys Rev E}\ }\textbf {\bibinfo {volume} {85}},\
  \bibinfo {pages} {016327} (\bibinfo {year} {2012})}\BibitemShut {NoStop}%
\bibitem [{\citenamefont {Rayleigh}(1884)}]{LordRayleigh1884}%
  \BibitemOpen
  \bibfield  {author} {\bibinfo {author} {\bibfnamefont {L.}~\bibnamefont
  {Rayleigh}},\ }\href@noop {} {\bibfield  {journal} {\bibinfo  {journal}
  {Philosophical Transactions of the Royal Society of London}\ }\textbf
  {\bibinfo {volume} {175}},\ \bibinfo {pages} {1} (\bibinfo {year}
  {1884})}\BibitemShut {NoStop}%
\bibitem [{\citenamefont {Muller}\ \emph {et~al.}(2012)\citenamefont {Muller},
  \citenamefont {Barnkob}, \citenamefont {Jensen},\ and\ \citenamefont
  {Bruus}}]{Muller2012}%
  \BibitemOpen
  \bibfield  {author} {\bibinfo {author} {\bibfnamefont {P.~B.}\ \bibnamefont
  {Muller}}, \bibinfo {author} {\bibfnamefont {R.}~\bibnamefont {Barnkob}},
  \bibinfo {author} {\bibfnamefont {M.~J.~H.}\ \bibnamefont {Jensen}}, \ and\
  \bibinfo {author} {\bibfnamefont {H.}~\bibnamefont {Bruus}},\ }\href
  {\doibase 10.1039/C2LC40612H} {\bibfield  {journal} {\bibinfo  {journal} {Lab
  Chip}\ }\textbf {\bibinfo {volume} {12}},\ \bibinfo {pages} {in press}
  (\bibinfo {year} {2012})}\BibitemShut {NoStop}%
\bibitem [{\citenamefont {Rednikov}\ and\ \citenamefont
  {Sadhal}(2011)}]{Rednikov2011}%
  \BibitemOpen
  \bibfield  {author} {\bibinfo {author} {\bibfnamefont {A.~Y.}\ \bibnamefont
  {Rednikov}}\ and\ \bibinfo {author} {\bibfnamefont {S.~S.}\ \bibnamefont
  {Sadhal}},\ }\href {\doibase 10.1017/S0022112010004532} {\bibfield  {journal}
  {\bibinfo  {journal} {Journal of Fluid Mechanics}\ }\textbf {\bibinfo
  {volume} {667}},\ \bibinfo {pages} {426} (\bibinfo {year}
  {2011})}\BibitemShut {NoStop}%
\bibitem [{\citenamefont {Fax{\'e}n}(1922)}]{Faxen1922}%
  \BibitemOpen
  \bibfield  {author} {\bibinfo {author} {\bibfnamefont {H.}~\bibnamefont
  {Fax{\'e}n}},\ }\href@noop {} {\bibfield  {journal} {\bibinfo  {journal} {Ann
  Phys}\ }\textbf {\bibinfo {volume} {68}},\ \bibinfo {pages} {89} (\bibinfo
  {year} {1922})}\BibitemShut {NoStop}%
\bibitem [{\citenamefont {Brenner}(1961)}]{Brenner1961}%
  \BibitemOpen
  \bibfield  {author} {\bibinfo {author} {\bibfnamefont {H.}~\bibnamefont
  {Brenner}},\ }\href {\doibase 10.1016/0009-2509(61)80035-3} {\bibfield
  {journal} {\bibinfo  {journal} {Chem Eng Sci}\ }\textbf {\bibinfo {volume}
  {16}},\ \bibinfo {pages} {242} (\bibinfo {year} {1961})}\BibitemShut
  {NoStop}%
\bibitem [{\citenamefont {Happel}\ and\ \citenamefont
  {Brenner}(1983)}]{Happel1983}%
  \BibitemOpen
  \bibfield  {author} {\bibinfo {author} {\bibfnamefont {J.}~\bibnamefont
  {Happel}}\ and\ \bibinfo {author} {\bibfnamefont {H.}~\bibnamefont
  {Brenner}},\ }\href@noop {} {\emph {\bibinfo {title} {Low Reynolds number
  hydrodynamics with special applications to particulate media}}}\ (\bibinfo
  {publisher} {Martinus Nijhoff Publishers},\ \bibinfo {address} {The Hague},\
  \bibinfo {year} {1983})\BibitemShut {NoStop}%
\bibitem [{\citenamefont {{CRCnetBASE Product}}(2012)}]{crc}%
  \BibitemOpen
  \bibfield  {author} {\bibinfo {author} {\bibnamefont {{CRCnetBASE
  Product}}},\ }\href@noop {} {\emph {\bibinfo {title} {CRC Handbook of
  Chemistry and Physics}}},\ \bibinfo {edition} {92nd}\ ed.\ (\bibinfo
  {publisher} {Taylor and Francis Group, www.hbcpnetbase.com/},\ \bibinfo
  {year} {2012})\BibitemShut {NoStop}%
\bibitem [{\citenamefont {Bergmann}(1954)}]{Bergmann1954}%
  \BibitemOpen
  \bibfield  {author} {\bibinfo {author} {\bibfnamefont {L.}~\bibnamefont
  {Bergmann}},\ }\href@noop {} {\emph {\bibinfo {title} {{Der Ultraschall und
  seine Anwendung in Wissenschaft und Technik}}}},\ \bibinfo {edition} {6th}\
  ed.\ (\bibinfo  {publisher} {S. Hirzel Verlag},\ \bibinfo {address}
  {Stuttgart},\ \bibinfo {year} {1954})\BibitemShut {NoStop}%
\bibitem [{\citenamefont {Mott}\ \emph {et~al.}(2008)\citenamefont {Mott},
  \citenamefont {Dorgan},\ and\ \citenamefont {Roland}}]{Mott2008}%
  \BibitemOpen
  \bibfield  {author} {\bibinfo {author} {\bibfnamefont {P.~H.}\ \bibnamefont
  {Mott}}, \bibinfo {author} {\bibfnamefont {J.~R.}\ \bibnamefont {Dorgan}}, \
  and\ \bibinfo {author} {\bibfnamefont {C.~M.}\ \bibnamefont {Roland}},\
  }\href {\doibase 10.1016/j.jsv.2008.01.026} {\bibfield  {journal} {\bibinfo
  {journal} {J Sound Vibr}\ }\textbf {\bibinfo {volume} {312}},\ \bibinfo
  {pages} {572} (\bibinfo {year} {2008})}\BibitemShut {NoStop}%
\bibitem [{\citenamefont {Landau}\ and\ \citenamefont
  {Lifshitz}(1986)}]{Landau1986}%
  \BibitemOpen
  \bibfield  {author} {\bibinfo {author} {\bibfnamefont {L.~D.}\ \bibnamefont
  {Landau}}\ and\ \bibinfo {author} {\bibfnamefont {E.~M.}\ \bibnamefont
  {Lifshitz}},\ }\href@noop {} {\emph {\bibinfo {title} {Theory of Elasticity.
  Course of Theoretical Physics}}},\ \bibinfo {edition} {3rd}\ ed.,\
  Vol.~\bibinfo {volume} {7}\ (\bibinfo  {publisher} {Pergamon Press},\
  \bibinfo {address} {Oxford},\ \bibinfo {year} {1986})\BibitemShut {NoStop}%
\bibitem [{\citenamefont {Cheng}(2008)}]{Cheng2008}%
  \BibitemOpen
  \bibfield  {author} {\bibinfo {author} {\bibfnamefont {N.-S.}\ \bibnamefont
  {Cheng}},\ }\href {\doibase 10.1021/ie071349z} {\bibfield  {journal}
  {\bibinfo  {journal} {Ind Eng Chem Res}\ }\textbf {\bibinfo {volume} {47}},\
  \bibinfo {pages} {3285} (\bibinfo {year} {2008})}\BibitemShut {NoStop}%
\bibitem [{\citenamefont {Fergusson}\ \emph {et~al.}(1954)\citenamefont
  {Fergusson}, \citenamefont {Guptill},\ and\ \citenamefont
  {MacDonald}}]{Fergusson1954}%
  \BibitemOpen
  \bibfield  {author} {\bibinfo {author} {\bibfnamefont {F.}~\bibnamefont
  {Fergusson}}, \bibinfo {author} {\bibfnamefont {E.}~\bibnamefont {Guptill}},
  \ and\ \bibinfo {author} {\bibfnamefont {A.}~\bibnamefont {MacDonald}},\
  }\href {\doibase 10.1121/1.1907292} {\bibfield  {journal} {\bibinfo
  {journal} {J Acoust Soc Am}\ }\textbf {\bibinfo {volume} {26}},\ \bibinfo
  {pages} {67} (\bibinfo {year} {1954})}\BibitemShut {NoStop}%
\bibitem [{\citenamefont {Mikkelsen}\ and\ \citenamefont
  {Bruus}(2005)}]{Mikkelsen2005}%
  \BibitemOpen
  \bibfield  {author} {\bibinfo {author} {\bibfnamefont {C.}~\bibnamefont
  {Mikkelsen}}\ and\ \bibinfo {author} {\bibfnamefont {H.}~\bibnamefont
  {Bruus}},\ }\href {\doibase 10.1039/b507104f} {\bibfield  {journal} {\bibinfo
   {journal} {Lab Chip}\ }\textbf {\bibinfo {volume} {5}},\ \bibinfo {pages}
  {1293} (\bibinfo {year} {2005})}\BibitemShut {NoStop}%
\bibitem [{\citenamefont {Meinhart}\ \emph {et~al.}(2000)\citenamefont
  {Meinhart}, \citenamefont {Wereley},\ and\ \citenamefont
  {Gray}}]{Meinhart2000}%
  \BibitemOpen
  \bibfield  {author} {\bibinfo {author} {\bibfnamefont {C.}~\bibnamefont
  {Meinhart}}, \bibinfo {author} {\bibfnamefont {S.}~\bibnamefont {Wereley}}, \
  and\ \bibinfo {author} {\bibfnamefont {M.}~\bibnamefont {Gray}},\ }\href@noop
  {} {\bibfield  {journal} {\bibinfo  {journal} {Measurement Science and
  Technology}\ }\textbf {\bibinfo {volume} {11}},\ \bibinfo {pages} {809}
  (\bibinfo {year} {2000})}\BibitemShut {NoStop}%
\bibitem [{\citenamefont {Olsen}\ and\ \citenamefont
  {Adrian}(2000)}]{Olsen2000}%
  \BibitemOpen
  \bibfield  {author} {\bibinfo {author} {\bibfnamefont {M.~G.}\ \bibnamefont
  {Olsen}}\ and\ \bibinfo {author} {\bibfnamefont {R.~J.}\ \bibnamefont
  {Adrian}},\ }\href@noop {} {\bibfield  {journal} {\bibinfo  {journal} {Exp
  Fluids}\ }\textbf {\bibinfo {volume} {Suppl.}},\ \bibinfo {pages} {S166}
  (\bibinfo {year} {2000})}\BibitemShut {NoStop}%
\bibitem [{\citenamefont {Rossi}\ \emph {et~al.}(2012)\citenamefont {Rossi},
  \citenamefont {Segura}, \citenamefont {Cierpka},\ and\ \citenamefont
  {Kaehler}}]{Rossi2012}%
  \BibitemOpen
  \bibfield  {author} {\bibinfo {author} {\bibfnamefont {M.}~\bibnamefont
  {Rossi}}, \bibinfo {author} {\bibfnamefont {R.}~\bibnamefont {Segura}},
  \bibinfo {author} {\bibfnamefont {C.}~\bibnamefont {Cierpka}}, \ and\
  \bibinfo {author} {\bibfnamefont {C.~J.}\ \bibnamefont {Kaehler}},\ }\href
  {\doibase 10.1007/s00348-011-1194-z} {\bibfield  {journal} {\bibinfo
  {journal} {Experiments in Fluids}\ }\textbf {\bibinfo {volume} {52}},\
  \bibinfo {pages} {1063} (\bibinfo {year} {2012})}\BibitemShut {NoStop}%
\bibitem [{\citenamefont {Adams}\ \emph {et~al.}(2012)\citenamefont {Adams},
  \citenamefont {Ebbesen}, \citenamefont {Barnkob}, \citenamefont {Yang},
  \citenamefont {Soh},\ and\ \citenamefont {Bruus}}]{Adams2012}%
  \BibitemOpen
  \bibfield  {author} {\bibinfo {author} {\bibfnamefont {J.~D.}\ \bibnamefont
  {Adams}}, \bibinfo {author} {\bibfnamefont {C.~L.}\ \bibnamefont {Ebbesen}},
  \bibinfo {author} {\bibfnamefont {R.}~\bibnamefont {Barnkob}}, \bibinfo
  {author} {\bibfnamefont {A.~H.~J.}\ \bibnamefont {Yang}}, \bibinfo {author}
  {\bibfnamefont {H.~T.}\ \bibnamefont {Soh}}, \ and\ \bibinfo {author}
  {\bibfnamefont {H.}~\bibnamefont {Bruus}},\ }\href {\doibase
  10.1088/0960-1317/22/7/075017} {\bibfield  {journal} {\bibinfo  {journal} {J
  Micromech Microeng}\ }\textbf {\bibinfo {volume} {22}},\ \bibinfo {pages}
  {075017} (\bibinfo {year} {2012})}\BibitemShut {NoStop}%
\bibitem [{\citenamefont {Yasuda}\ and\ \citenamefont
  {Kamakura}(1997)}]{Yasuda1997}%
  \BibitemOpen
  \bibfield  {author} {\bibinfo {author} {\bibfnamefont {K.}~\bibnamefont
  {Yasuda}}\ and\ \bibinfo {author} {\bibfnamefont {T.}~\bibnamefont
  {Kamakura}},\ }\href {http://link.aip.org/link/?APPLAB/71/1771/1} {\bibfield
  {journal} {\bibinfo  {journal} {Applied physics letters}\ }\textbf {\bibinfo
  {volume} {71}},\ \bibinfo {pages} {1771} (\bibinfo {year}
  {1997})}\BibitemShut {NoStop}%
\bibitem [{\citenamefont {Hags\"{a}ter}\ \emph {et~al.}(2008)\citenamefont
  {Hags\"{a}ter}, \citenamefont {Lenshof}, \citenamefont {{Skafte-Pedersen}},
  \citenamefont {Kutter}, \citenamefont {Laurell},\ and\ \citenamefont
  {Bruus}}]{Hagsater2008}%
  \BibitemOpen
  \bibfield  {author} {\bibinfo {author} {\bibfnamefont {S.~M.}\ \bibnamefont
  {Hags\"{a}ter}}, \bibinfo {author} {\bibfnamefont {A.}~\bibnamefont
  {Lenshof}}, \bibinfo {author} {\bibfnamefont {P.}~\bibnamefont
  {{Skafte-Pedersen}}}, \bibinfo {author} {\bibfnamefont {J.~P.}\ \bibnamefont
  {Kutter}}, \bibinfo {author} {\bibfnamefont {T.}~\bibnamefont {Laurell}}, \
  and\ \bibinfo {author} {\bibfnamefont {H.}~\bibnamefont {Bruus}},\ }\href
  {\doibase 10.1039/b801028e} {\bibfield  {journal} {\bibinfo  {journal} {Lab
  Chip}\ }\textbf {\bibinfo {volume} {8}},\ \bibinfo {pages} {1178} (\bibinfo
  {year} {2008})}\BibitemShut {NoStop}%
\end{thebibliography}
\end{document}